\documentclass{article}
\usepackage[a4paper, margin=1in]{geometry}
\usepackage[parfill]{parskip}
\usepackage{hyperref}
\usepackage{url}
\usepackage{natbib}
\usepackage{amsmath}
\usepackage{amssymb}
\usepackage{amsthm}
\usepackage{amsfonts}
\usepackage{mathtools}
\usepackage{thmtools}
\usepackage{xcolor}
\usepackage{graphicx}
\usepackage{tikz}
\usetikzlibrary{arrows.meta,calc,patterns,3d}
\usepackage{pgfplots}
\pgfplotsset{compat=1.18}
\usepackage{subcaption}

\usepackage{cleveref}
\Crefname{subfigure}{Figure}{Figures}
\usepackage{authblk}
\usepackage{algorithm}
\usepackage[noend]{algpseudocode}
\algrenewcommand\algorithmicrequire{\textbf{Input:}}
\algrenewcommand\algorithmicensure{\textbf{Output:}}

\newtheorem{theorem}{Theorem}[section]
\newtheorem{corollary}[theorem]{Corollary}
\newtheorem{lemma}[theorem]{Lemma}
\newtheorem{proposition}[theorem]{Proposition}

\theoremstyle{definition}
\newtheorem{definition}[theorem]{Definition}

\newcommand{\Newt}{\operatorname{Newt}}
\newcommand{\conv}{\operatorname{conv}}

\title{
Parameterized Hardness of Zonotope Containment\\and Neural Network Verification
\thanks{This article is an extension of a preliminary conference version that appeared in the proceedings of the ICLR 2026 conference \citep{froese2026parameterized}.}
}

\author[1]{Vincent Froese}
\author[2]{Moritz Grillo}
\author[3]{Christoph Hertrich}
\author[3]{Moritz Stargalla}
\affil[1]{\small Technische Universität Berlin}
\affil[2]{\small Max Planck Institute for Mathematics in the Sciences}
\affil[3]{\small University of Technology Nuremberg}

\date{September 3, 2026}

\newcommand{\problemdef}[3]{%
  \begin{center}
    \begin{minipage}{0.95\linewidth}
      \noindent
      \textsc{#1}
      \vspace{5pt}\\
      \setlength{\tabcolsep}{3pt}%
      \begin{tabular}{@{}l p{0.85\linewidth}@{}}
        \textbf{Input:}    & #2 \\
        \textbf{Question:} & #3
      \end{tabular}
    \end{minipage}
  \end{center}
}

\definecolor{OIblue}{RGB}{86,180,233}
\definecolor{OIorange}{RGB}{213,94,0}
\definecolor{OIgreen}{RGB}{0,158,115}

\tikzset{
	relu pic/.pic={
		\draw[thick] (-0.11,0) -- (0,0) -- (0.09,0.12);
	}
}

\newcommand{\R}{\mathbb{R}}
\newcommand{\N}{\mathbb{N}}
\DeclareMathOperator{\poly}{poly}

\newsavebox{\leftbox}
\newsavebox{\rightbox}
\newlength{\contentH}

\sbox{\leftbox}{
  \begin{minipage}{0.55\linewidth}
    \includegraphics[width=\linewidth, page=1]{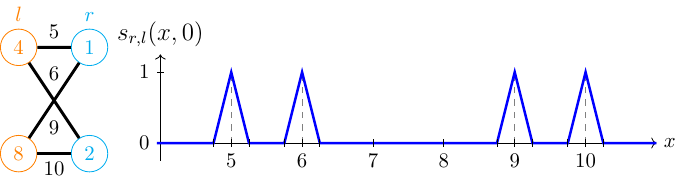}
    \includegraphics[width=\linewidth, page=2]{tikz_figures.pdf}\vspace*{0.075cm}
  \end{minipage}
}
\sbox{\rightbox}{
  \includegraphics[width=.44\linewidth, page=3]{tikz_figures.pdf}
}

\ifdim \dimexpr\ht\rightbox+\dp\rightbox\relax > \contentH
\fi

\begin{document}

\maketitle

\begin{abstract}
\noindent
Neural networks with ReLU activations are a widely used model in machine learning. It is thus important to have a profound understanding of the properties of the functions computed by such networks. Recently, there has been increasing interest in the (parameterized) computational complexity of determining these properties. In this work, we close several gaps and resolve an open problem posed by Froese~et~al.~[COLT~'25] regarding the parameterized complexity of various problems related to network verification.

In particular, we prove that, for all~$\ell\ge 2$, deciding positivity (and thus surjectivity) of a function $f\colon\R^d\to\R$ computed by an $\ell$-layer ReLU network is W[$\ell-1$]-hard when parameterized by the input dimension~$d$.
The case $\ell=2$ implies that zonotope non-containment (a problem that is of independent interest in computational geometry, control theory, and robotics) is W[1]-hard with respect to the ambient dimension~$d$.
Moreover, we show that approximating the maximum within any multiplicative factor and computing the~$L_p$-Lipschitz constant for~$p\in(0,\infty]$ in $\ell$-layer networks is NP-hard and W[$\ell-1$]-hard with respect to~$d$. For~$\ell\ge 3$, approximating the~$L_p$-Lipschitz constant is NP- and~W[$\ell-2$]-hard.
We further show that the above problems are NP- and~W[$t$]-hard (for all~$t\ge 1$) with respect to~$\ell$ for constant~$d$.

Notably, our hardness results imply that the naive enumeration-based methods for these fundamental problems running in $n^{(\ell-1) d}\cdot\poly(N)$ time are all essentially optimal under the Exponential Time Hypothesis.
\end{abstract}

\section{Introduction}
Neural networks with rectified linear unit (ReLU) activations are a common model in deep learning. In practice, such networks are trained on finite datasets and are expected to generalize reliably to unseen inputs. However, even minor perturbations of the input may lead to unexpected or erroneous outputs~\citep{SzegedyZSBEGF13}. This highlights the importance of explaining and certifying trained models, which in turn requires a detailed understanding of the functions computed by ReLU networks.

A central problem in this context is \emph{network verification}: Given a set~$\mathcal X$ of inputs, the question is whether the network’s outputs are guaranteed to lie within a prescribed set~$\mathcal Y$. Commonly, the sets~$\mathcal X$ and~$\mathcal Y$ take the form of balls or are specified by linear constraints. This question has received increasing attention in recent years, particularly due to the deployment of neural networks in safety-critical applications \citep{bojarski2016endendlearningselfdriving,Weng18,KL21,RP21,KBDJK22}.
Recently, \citet{FGS25} established a connection between the basic verification task to decide \emph{positivity} (that is, whether a 2-layer ReLU network attains a positive output) and the classical geometry problem of \emph{zonotope containment}. The latter asks whether one zonotope is contained within another, a question that has been extensively studied due to its applications in areas such as robotics and control~\citep{sadraddini2019linear,gruber2020computing,gruber2021scalable,KULMBURG202184,yang2022efficient,KSA25,ERRS26}.

Beyond verification, robustness is often a crucial requirement since trained networks are typically expected to be insensitive to small input perturbations. This property is commonly quantified in terms of the network’s \emph{Lipschitz constant}, which should ideally be small \citep{VS18,weng2018evaluating,fazlyab2019efficient,JD20,bhowmick2021lipbab,splittgerber2026exlipbab}.

Network verification \citep{KBDJK22,salzer2023,FGS25}, estimating the Lipschitz constant \citep{VS18,JD20} and zonotope containment \citep{KULMBURG202184} are all known to be (co)NP-hard.
This intractability is closely linked to the ``curse of dimensionality'' and the ``curse of depth''. With growing input dimension or growing number of layers, the search space becomes prohibitively large:
The number of linear regions of an $\ell$-layer network with at most~$n$ neurons per hidden layer is in~$\mathcal O(n^{(\ell-1) d})$.
Hence, the naive approach to enumerate all these regions quickly becomes intractable.

A natural question is whether the problems become tractable for low-dimensional input spaces or shallow networks.
This perspective is particularly relevant since, in practice, high-dimensional data is often assumed to lie near a low-dimensional submanifold of the input space.
In addition, shallow networks can be trained more efficiently in practice and yield simpler models.
Motivated by this, recent work has studied the \emph{parameterized complexity} of neural network problems such as training \citep{AroraBMM18,FHN22,brand2023new,FH23,GSS26}, verification \citep{FGS25} and interpretability \citep{BM0S20,AVW25} (see also \citep{ganian2026parameterized} for a survey).
\textcolor{black}{Notably, while checking injectivity of a 2-layer ReLU network with~$n$ hidden neurons can be done in $(d+1)^d\cdot n^{\mathcal{O}(1)}$ time (that is, it is \emph{fixed-parameter tractable} with respect to~$d$) \citep{FGS25}, the parameterized complexity status of network verification (in particular the positivity question) and the Lipschitz constant have been posed as open problems at COLT~'25 \citep{FGHS25}.}

\subsection{Our Contributions}
We answer the aforementioned questions by proving several W-hardness results for the parameter~$d$ and also for~$\ell$, thus excluding fixed-parameter tractability under standard complexity assumptions (even if the other parameter is a constant).
Moreover, we show that solving these problems via simple ``brute-force'' enumeration of the linear regions of the network's function is essentially optimal under the Exponential Time Hypothesis (ETH). 
(For a visual overview of our results, see \Cref{tab:param-complexity-overview-two-parameters}.)

In \Cref{sec:gadget_construction}, we give a reduction from the well-known \textsc{Multicolored Clique} problem to positivity for 2-layer networks in which the network's input dimension depends linearly on the clique size.
This reduction forms the basis for our hardness results and yields strong running time lower bounds based on the ETH.
The key difficulty here is that the input dimension must scale linearly with the clique size (in contrast, standard NP-hardness reductions allow the input dimension to grow without restriction).

\paragraph{Network Verification.}
We study the NP-hard problems of deciding positivity, deciding surjectivity, and approximating the maximum of a ReLU network $f\colon\R^d\to\R$ with $\ell$ layers,~$n$ hidden neurons and encoding size $N$.
All these problems are complements of special verification tasks.
For example, positivity corresponds to checking whether there exists $x\in\R^d$ with $f(x)>0$, that is, $f(\R^d)\not \subseteq  (-\infty,0]$.
The problems can all be solved in $n^{\mathcal{O}((\ell-1) d)}\cdot \poly(N)$ time with simple ``brute-force'' enumeration algorithms (see \Cref{sec:algorithms_and_containment}).
In \Cref{sec:verification}, we prove W[$\ell-1$]-hardness with respect to~$d$ for all problems with constant~$\ell\ge 2$. Thus, increasing the depth leads to hardness for higher levels of the W-hierarchy.
The case $\ell=2$ resolves the open question by~\citet{FGHS25}.
Moreover, our reductions imply a running time lower bound of $n^{\Omega(d)}\cdot \poly(N)$ for constant~$\ell$ based on the ETH which shows that the simple enumeration algorithms are essentially optimal.
In \Cref{sec:paraNP}, we show NP-hardness and hardness for every finite level of the W-hierarchy for all problems when parameterized by~$\ell$ even for constant $d$. Again, this implies a lower bound of $n^{\Omega(\ell)}\cdot \poly(N)$ for fixed~$d$.

Clearly, all running time lower bounds also hold for the general network verification problem.
That is, in \Cref{sec:paraNP}, we show that the running time $n^{(\ell-1) d}\cdot\poly(N)$ for network verification is essentially optimal.

\paragraph{Zonotope Containment.}
In \Cref{sec:zonotope_containment_hardness}, we study the coNP-hard problem of deciding whether a zonotope $Z\subset \R^d$ (given by its generators) is contained in another zonotope $Z'\subset \R^d$.
Based on a duality of 2-layer ReLU networks and zonotopes, we obtain coW[1]-hardness with respect to~$d$ and an analogous running time lower bound of $n^{\Omega(d)}\cdot \poly(N)$ assuming the ETH which shows that the vertex enumeration algorithm is essentially optimal.

\paragraph{Lipschitz Constant.}
\citet{VS18} proved that computing the $L_2$-Lipschitz constant of a 2-layer ReLU network is NP-hard.
In \Cref{sec:lipschitz_hardness}, we extend this to NP-hardness for every $p\in(0,\infty]$ and even W[1]-hardness with respect to~$d$.
Approximating the $L_p$-Lipschitz constant within any multiplicative constant for 3-layer ReLU networks is known to be NP-hard \citep{JD20,FGS25}.
We also extend this result to W[1]-hardness with respect to~$d$.
Again, our reductions imply running time lower bounds matching the running times of simple enumeration algorithms.
Moreover, we extend the W[1]-hardness for the 2-layer (3-layer) version to W[$\ell$]-hardness for the $(\ell+1)$-layer ($(\ell+2)$-layer) version.
In \Cref{sec:paraNP}, we also obtain the same hardness results regarding the parameter~$\ell$ for constant~$d$ as for the network verification tasks.
On the positive side, we show that for the restricted class of \emph{input convex} networks, computing the $L_1$-Lipschitz constant is polynomial-time solvable and the $L_\infty$-Lipschitz constant is \emph{fixed-parameter tractable} (FPT) with respect to~$d$.
In \Cref{sec:norm_max_fpt_approx}, we discuss the equivalence between Lipschitz constant computation and norm maximization on zonotopes and present a randomized FPT-approximation algorithm, using results from subspace embeddings.
Recently, several independent works showed that computing the $L_p$-Lipschitz constant (and $L_p$-norm maximization over zonotopes) is W[1]-hard with respect to $d$ for every $p\in (1,\infty)$ \citep{cao2026ellpnorm,dewasurendra2026convex,das2026parameterized}.
Moreover, \citet{cao2026ellpnorm} also present a deterministic FPT-approximation algorithm.

\paragraph{Injectivity.}
\cite{FGS25} showed that deciding whether a ReLU network with a single hidden layer computes an injective map is coNP-complete, but the problem is FPT when parameterized by the input dimension $d$. We show that the fixed-parameter tractability is a property of shallow networks by proving coW[$\ell-2$]-hardness for $\ell \ge 3$ layers (\Cref{thm:injectivity_Whardness}) thereby answering the open question of \citet{FGS25}. 

From a wider perspective, besides settling the parameterized complexity status of several important ReLU neural network (and zonotope) problems, our work establishes an interesting connection between deep neural networks and parameterized complexity theory by relating the W-hierarchy which is based on bounded-weft Boolean circuits to bounded-depth ReLU neural networks.
This might lead to further interesting insights connecting these two (so far rather unrelated) research areas.

\definecolor{headergray}{RGB}{248,248,248}
\definecolor{bandgray}{RGB}{238,238,238}
\definecolor{naGray}{RGB}{232,232,232}
\definecolor{containmentcol}{RGB}{204,235,230}
\definecolor{hardnesscol}{RGB}{253,218,198}
\definecolor{ethcol}{RGB}{214,228,245}

\def\TallHardTop{0.755}
\def\TallHardBottom{0.27}

\def\ThmTagW{0.44}
\def\ThmTagH{0.20}
\newcommand{\ThmTagFont}{\fontsize{4.8}{5.0}\selectfont}

\newcommand{\ThmTag}[3]{%
    \if\relax\detokenize{#3}\relax
    \else
        \draw[line width=0.25pt]
            ({#1-\ThmTagW},{#2+\ThmTagH}) -- (#1,{#2+\ThmTagH});
        \draw[line width=0.25pt]
            ({#1-\ThmTagW},#2) -- ({#1-\ThmTagW},{#2+\ThmTagH});
        \node[font=\ThmTagFont, inner sep=0pt]
            at ({#1-\ThmTagW/2},{#2+\ThmTagH/2}) {\ref{#3}};
    \fi
}

\newcommand{\ResultCell}[8]{%
\begin{tikzpicture}[x=1cm,y=1cm,baseline=(current bounding box.center)]
    \def\w{#1}\def\h{#2}
    \fill[containmentcol] (0,{2*\h/3}) rectangle (\w,\h);
    \fill[hardnesscol]   (0,{\h/3}) rectangle (\w,{2*\h/3});
    \fill[ethcol]        (0,0) rectangle (\w,{\h/3});
    \draw[line width=0.35pt] (0,0) rectangle (\w,\h);
    \draw[line width=0.25pt] (0,{\h/3}) -- (\w,{\h/3});
    \draw[line width=0.25pt] (0,{2*\h/3}) -- (\w,{2*\h/3});

    \node[font=\scriptsize, align=center, inner sep=1pt, text width=#1cm]
        at ({\w/2},{5*\h/6}) {#3};
    \node[font=\scriptsize, align=center, inner sep=1pt, text width=#1cm]
        at ({\w/2},{\h/2}) {#4};
    \node[font=\scriptsize, align=center, inner sep=1pt, text width=#1cm]
        at ({\w/2},{\h/6}) {#5};

    \ThmTag{\w}{2*\h/3}{#6}
    \ThmTag{\w}{\h/3}{#7}
    \ThmTag{\w}{0}{#8}
\end{tikzpicture}%
}

\newcommand{\ResultCellTallHard}[8]{%
\begin{tikzpicture}[x=1cm,y=1cm,baseline=(current bounding box.center)]
    \def\w{#1}\def\h{#2}
    \fill[containmentcol] (0,{\TallHardTop*\h}) rectangle (\w,\h);
    \fill[hardnesscol]   (0,{\TallHardBottom*\h}) rectangle (\w,{\TallHardTop*\h});
    \fill[ethcol]        (0,0) rectangle (\w,{\TallHardBottom*\h});
    \draw[line width=0.35pt] (0,0) rectangle (\w,\h);
    \draw[line width=0.25pt] (0,{\TallHardBottom*\h}) -- (\w,{\TallHardBottom*\h});
    \draw[line width=0.25pt] (0,{\TallHardTop*\h}) -- (\w,{\TallHardTop*\h});

    \node[font=\scriptsize, align=center, inner sep=1pt, text width=#1cm]
        at ({\w/2},{(1+\TallHardTop)*\h/2}) {#3};
    \node[font=\scriptsize, align=center, inner sep=1pt, text width=#1cm]
        at ({\w/2},{(\TallHardTop+\TallHardBottom)*\h/2}) {#4};
    \node[font=\scriptsize, align=center, inner sep=1pt, text width=#1cm]
        at ({\w/2},{\TallHardBottom*\h/2}) {#5};

    \ThmTag{\w}{\TallHardTop*\h}{#6}
    \ThmTag{\w}{\TallHardBottom*\h}{#7}
    \ThmTag{\w}{0}{#8}
\end{tikzpicture}%
}

\begin{table}[t]
\centering
\resizebox{.9\textwidth}{!}{%
\begin{tikzpicture}[x=1cm,y=1cm]

\def\wProb{4.0}
\def\wD{4.8}
\def\wEll{4.8}
\def\hHead{0.90}
\def\hBand{0.72}
\def\hRow{1.60}
\def\wCell{4.35}
\def\hCell{1.26}

\pgfmathsetmacro{\xProb}{\wProb}
\pgfmathsetmacro{\xD}{\wProb+\wD}
\pgfmathsetmacro{\xEnd}{\wProb+\wD+\wEll}

\pgfmathsetmacro{\yHeadBottom}{-\hHead}
\pgfmathsetmacro{\yRone}{-(\hHead+\hRow)}
\pgfmathsetmacro{\yRtwo}{-(\hHead+2*\hRow)}
\pgfmathsetmacro{\yRthree}{-(\hHead+3*\hRow)}
\pgfmathsetmacro{\yRfour}{-(\hHead+4*\hRow)}
\pgfmathsetmacro{\yBottom}{-(\hHead+5*\hRow)}

\pgfmathsetmacro{\xProbC}{\wProb/2}
\pgfmathsetmacro{\xDC}{\wProb+\wD/2}
\pgfmathsetmacro{\xEllC}{\wProb+\wD+\wEll/2}

\pgfmathsetmacro{\yHeadC}{-\hHead/2}
\pgfmathsetmacro{\yOne}{-(\hHead+0.5*\hRow)}
\pgfmathsetmacro{\yTwo}{-(\hHead+1.5*\hRow)}
\pgfmathsetmacro{\yThree}{-(\hHead+2.5*\hRow)}
\pgfmathsetmacro{\yFour}{-(\hHead+3.5*\hRow)}
\pgfmathsetmacro{\yFive}{-(\hHead+4.5*\hRow)}

\fill[headergray] (0,0) rectangle (\xEnd,\yHeadBottom);
\fill[naGray] (\xD,\yHeadBottom) rectangle (\xEnd,\yRtwo);

\draw[line width=0.55pt] (0,0) rectangle (\xEnd,\yBottom);

\draw[line width=0.45pt] (\xProb,0) -- (\xProb,\yBottom);
\draw[line width=0.45pt] (\xD,0) -- (\xD,\yBottom);

\draw[line width=0.45pt] (0,\yHeadBottom) -- (\xEnd,\yHeadBottom);
\draw[line width=0.40pt] (0,\yRone) -- (\xD,\yRone);
\draw[line width=0.45pt] (0,\yRtwo) -- (\xEnd,\yRtwo);
\draw[line width=0.40pt] (0,\yRthree) -- (\xEnd,\yRthree);
\draw[line width=0.40pt] (0,\yRfour) -- (\xEnd,\yRfour);

\draw[line width=0.45pt] (\xD,\yHeadBottom) -- (\xEnd,\yRtwo);
\draw[line width=0.45pt] (\xD,\yRtwo) -- (\xEnd,\yHeadBottom);

\node[font=\normalsize\bfseries] at (\xProbC,\yHeadC) {Problem};
\node[font=\normalsize\bfseries] at (\xDC,\yHeadC) {Parameter $d$};
\node[font=\normalsize\bfseries] at (\xEllC,\yHeadC) {Parameter $\ell$};

\node[anchor=west, align=left, font=\small, text width=3.6cm]
    at (0.18,\yOne)
    {$L_p$-Max on Zonotopes\\\begin{center}$(p\in (0,\infty])$\end{center}};

\node[anchor=west, align=left, font=\small, text width=3.6cm]
    at (0.18,\yTwo)
    {Zonotope\\ Non-Containment};

\node[anchor=west, align=left, font=\small, text width=3.6cm]
    at (0.18,\yThree)
    {Positivity};

\node[anchor=west, align=left, font=\small, text width=3.6cm]
    at (0.18,\yFour)
    {Surjectivity};

\node[anchor=west, align=left, font=\small, text width=3.6cm]
    at (0.18,\yFive)
    {$L_p$-Lipschitz Constant\\\begin{center}$(p\in (0,\infty])$\end{center}};

\node at (\xDC,\yOne)
    {\ResultCellTallHard{\wCell}{\hCell}
        {W[P]}
        {
        W[1]-hard for $p\in (1,\infty)\cap \mathbb{Q}$
        
        {\tiny\citep{cao2026ellpnorm,das2026parameterized}}
        }
        {$N^{\Theta(d)}$ for $p\in (1,\infty)\cap \mathbb{Q}$}
        {thm:WP-containment}
        {}
        {}};

\node at (\xDC,\yTwo)
    {\ResultCell{\wCell}{\hCell}
        {W[P]}
        {W[1]-hard}
        {$N^{\Theta(d)}$}
        {thm:WP-containment}
        {thm:zonotope-cont-hardness}
        {}};

\node at (\xDC,\yThree)
    {\ResultCell{\wCell}{\hCell}
        {W[P] for $\ell=2$}
        {W[$\ell-$1]-hard for all $\ell\ge 2$}
        {$N^{\Theta(d)}$}
        {thm:WP-containment}
        {thm:Wl-hardness}
        {}};

\node at (\xDC,\yFour)
    {\ResultCell{\wCell}{\hCell}
        {W[P] for $\ell=2$}
        {W[$\ell-$1]-hard for all $\ell\ge 2$}
        {$N^{\Theta(d)}$}
        {thm:WP-containment}
        {thm:surjectivity_Whardness}
        {}};

\node at (\xDC,\yFive)
    {\ResultCell{\wCell}{\hCell}
        {W[P] for $\ell=2$}
        {W[$\ell-$1]-hard for all $\ell\ge 2$}
        {$N^{\Theta(d)}$}
        {thm:WP-containment}
        {thm:Wl-hardness-Lipschitz}
        {}};


\node at (\xEllC,\yThree)
    {\ResultCellTallHard{\wCell}{\hCell}
        {W[P] for $d=1$}
        {$d=1$: NP-hard and W[$t$]-hard for all $t\ge 1$}
        {$N^{\Theta(\ell)}$}
        {thm:WP-containment_param_layers}
        {thm:fixed_d=1_hardness_positivity}
        {}};

\node at (\xEllC,\yFour)
    {\ResultCellTallHard{\wCell}{\hCell}
        {P for $d=1$}
        {$d=2$: NP-hard and W[$t$]-hard for all $t\ge 1$}
        {$N^{\Theta(\ell)}$}
        {thm:WP-containment_param_layers}
        {cor:fixed_d=2_hardness_surjectivity}
        {}};

\node at (\xEllC,\yFive)
    {\ResultCellTallHard{\wCell}{\hCell}
        {W[P] for $d=1$}
        {$d=1$: NP-hard and W[$t$]-hard for all $t\ge 1$}
        {$N^{\Theta(\ell)}$}
        {thm:WP-containment_param_layers}
        {cor:fixed_d=1_hardness_approx_lipschitz}
        {}};

\end{tikzpicture}%
}
\caption{
Parameterized complexity results for (most of) the considered problems of this paper with their theorem number (bottom right corner).
Green cells denote containment results, orange cells denote hardness results, and blue cells denote the optimal running time that is sufficient and necessary (assuming the ETH) to solve the problem.
For the ReLU network problems, $d$ denotes the input dimension and $\ell$ the number of layers of the networks.
For the zonotope problems, $d$ denotes the dimension of the ambient vector space.
$N$ denotes the encoding size of the input.
By \Cref{thm:enumeration_algorithms}, all ReLU problems are in XP with respect to the combined parameter $\ell d$.
}
\label{tab:param-complexity-overview-two-parameters}
\end{table}

\subsection{Further Related Work}
Various heuristic methods for network verification have been proposed, including interval bound propagation \citep{Gowal2018OnTE}, DeepZ \citep{Wong}, DeepPoly \citep{Singh}, multi-neuron verification~\cite{ferrari2022complete}, ZonoDual~\citep{jordan2022_zonotope}, and cutting planes \citep{Cuttingplane}.
\textcolor{black}{\citet{baader2024expressivity} and \citet{mao2026expressiveness} study the expressivity of convex relaxations that are often used in practical network verification algorithms.}
\citet{HP24} showed that ReLU neural networks (with unbounded weight precision) are strictly more powerful than Boolean circuits.
$L_p$-norm maximization on zonotopes (a special case of the Lipschitz constant problem) is also known as the \textsc{Longest Vector Sum} problem and has a wide range of applications in pattern recognition, clustering, signal processing, and analysis of large-scale data~\citep{BP07,Shenmaier18,Shenmaier20}.
Special cases were studied before~\citep{BGKL90}. The $L_2$-case corresponds to \textsc{Unconstrained Quadratic Binary Maximization} with positive semidefinite matrices of rank~$d$~\citep{FFL05}.

\section{Preliminaries}
\label{sec:prelims}

\paragraph*{Notation.}
For $n\in\N$, we define $[n]\coloneqq\{1,\ldots,n\}$.
For $k,n\in \N, k\leq n$, we define $\binom{[n]}{k}\coloneqq \{A\subseteq [n]: |A|=k\}$.
A function $f\colon\R^d\to \R^m$ is \emph{positively homogeneous} if $f(\lambda x)=\lambda f(x)$ holds for all $x\in \R^d$ and $\lambda \geq 0$.
Given a \emph{generator} matrix $A=(a_1,\dots, a_n)\in \R^{d\times n}$, the corresponding \emph{zonotope} is $Z(A)\coloneqq\sum_{i=1}^n \conv(\{0,a_i\})$, where the sum is the Minkowski sum of the generators.

\paragraph{$L_p$-Lipschitz Constant.}
For $p\in(0,\infty)$ and a vector $x\in \R^d$, we define $\|x\|_p\coloneqq\left(\sum_{i=1}^d |x_i|^p\right)^{\frac{1}{p}}$, and for $p=\infty$ we set $\|x\|_\infty\coloneqq\max_{i\in [d]}|x_i|$.
For $p\in [1,\infty]$, the function $\|\cdot\|_p$ is the \emph{$L_p$-norm}, and for $p\in(0,1)$, it is the \emph{$L_p$-quasinorm}.
The $L_p$-Lipschitz constant of a function $f$ is $L_p(f)\coloneqq\sup_{x\neq y} \frac{\|f(x)-f(y)\|_p}{\|x-y\|_p}$.

\paragraph{ReLU Networks.}
A \emph{ReLU layer} with $d$ inputs, $m$ outputs, weights $W\in \R^{m\times d}$, and biases $b\in \R^m$ computes the map $\phi_{W,b}\colon\R^d\to \R^m,\; x\mapsto \max(0, Wx+b)$, where the maximum is applied in each component.
A \emph{ReLU network} with $\ell\geq 1$ layers and one-dimensional output is defined by $\ell$ weight matrices~$W_i\in\R^{n_i\times n_{i-1}}$ and biases~$b_i\in\R^{n_i}$ for $i\in[\ell]$, where $n_0\coloneqq d,\ldots,n_\ell\coloneqq 1\in\N^+$, and computes the \emph{continuous piecewise linear} (CPWL) function $f\colon\R^d\to \R$ with
\[
f(x)\coloneqq W_\ell\cdot (\phi_{W_{\ell-1},b_{\ell-1}}\circ \cdots \circ \phi_{W_1,b_1})(x)+b_{\ell}.
\]
Observe that no activation function is applied in the output layer. The $\ell-1$ ReLU layers are also called \emph{hidden layers}.
The \emph{width} of the network is $\max\{n_1,\dots,n_{\ell-1}\}$ and its \emph{size} is $\sum_{i=1}^{\ell-1}n_i$.
The \emph{preactivation} of a neuron is the function computed by the neuron before applying the ReLU function.
A neuron is \emph{active} on a point if its preactivation is $>0$ on that particular point.

\paragraph{Geometry of ReLU Networks.}
We repeat basic definitions from polyhedral geometry, see \citet{ilp_theory} for more details.
A \emph{polyhedron} $P$ is the intersection of finitely many closed halfspaces.
A \emph{polyhedral cone} $C\subseteq \R^d$ is a polyhedron such that $\lambda u + \mu v \in C$ for every $u,v\in C$ and $\lambda, \mu \in \R_{\geq 0}$.
A cone is \emph{pointed} if it does not contain a straight line.
A \emph{ray} $\rho$ is a one-dimensional pointed cone; a vector $r$ is a \emph{ray generator} of $\rho$ if $\rho=\{\lambda r: \lambda \geq 0\}$.
A \emph{polyhedral complex} $\mathcal{P}$ is a finite collection of polyhedra such that $\emptyset \in \mathcal{P}$, if $P\in \mathcal{P}$, then all faces of $P$ are in $\mathcal{P}$, and if $P,P'\in \mathcal{P}$, then $P\cap P'$ is a face of $P$ and $P'$.
A \emph{cell} is a full-dimensional element of a polyhedral complex.
A \emph{hyperplane arrangement} $\mathcal{H}$ is a finite collection of hyperplanes in~$\R^d$.

A \emph{ReLU network} computing the CPWL function $f\colon\R^d\to \R$ with
\[
f(x)\coloneqq W_\ell\cdot (\phi_{W_{\ell-1},b_{\ell-1}}\circ \cdots \circ \phi_{W_1,b_1})(x)+b_{\ell},
\]
is affine linear on each cell of an associated polyhedral complex $\Sigma_f$ (the one induced by the closure of the full-dimensional activation regions \citep{hanin2019_deep} of the network).
In particular, the encoding size of every polyhedron in this complex is polynomially bounded in the encoding size of the neural network \citep{FGS25}.
It is well known that for 2-layer ReLU networks, $\Sigma_f$ corresponds to a hyperplane arrangement \citep{montufar2014number}.

\paragraph{Hyperplane Arrangements.}
Any hyperplane arrangement with $n$ hyperplanes in $d$ dimensions has at most $\mathcal{O}(n^d)$ many cells \citep{Zas75}, and it is possible to enumerate them in $\mathcal{O}(n^d)$ time \citep{edelsbrunner1986constructing}.
All zero- and one-dimensional faces of a hyperplane arrangement can be enumerated in $n^d\cdot \poly(N)$ and $n^{d-1}\cdot \poly(N)$ time, respectively, where $N$ is the encoding size of the hyperplane arrangement, since they arise from an intersection of $d$ and $d-1$ hyperplanes.

\paragraph{Maxout Networks.}
A \emph{2-maxout layer} with $d$ inputs, $m$ outputs, weight matrices $W^{(1)},W^{(2)}\in\R^{m\times d}$, and two bias vectors $b^{(1)},b^{(2)}\in \R^m$ computes the map
\[\psi_{W^{(1)},W^{(2)},b^{(1)},b^{(2)}}\colon\R^d\to \R^m,\quad x\mapsto \max(W^{(1)}x+b^{(1)}, W^{(2)}x+b^{(2)}),\]
where the maximum is applied in each component.
A \emph{2-maxout network} is defined analogously to a ReLU network, replacing the ReLU layers $\phi_{W_{i},b_{i}}$ with 2-maxout layers $\psi_{W^{(1)}_i,W^{(2)}_i,b^{(1)}_i,b^{(2)}_i}$ for all $i\in[\ell-1]$.
We note that ReLU networks form a subclass of maxout networks by choosing $W^{(1)}_i=0$ and $b^{(1)}_i=0$ for all $i\in [\ell-1]$.
In the general setting, maxout networks are essentially equivalent to ReLU networks (every $\ell$-layer 2-maxout network can be simulated by an $\ell$-layer ReLU network whose width is at most three times the width of the maxout network).
However, restricted to input convex neural networks (ICNNs), maxout networks are strictly more expressive than ReLU networks \citep{bakaev2025depth}.
This motivates studying maxout ICNNs as a generalization of ReLU ICNNs in \Cref{subsection:icnns}.

\paragraph*{Polytopes and Duality.}
 There is a duality between positively homogeneous convex CPWL functions from $\R^d$ to $\R$ (the set of which is denoted~$\mathcal{F}_d$) and polytopes in $\R^d$ (denoted~$\mathcal{P}_d$), which we will briefly sketch here; compare also \citet{HBDS23}.
 Any function $f \in \mathcal{F}_d$ can be written as $f(x) = \max\{a_1^\top x, \dots, a_k^\top x\}$ for some $a_i \in \R^d$, and its \emph{Newton polytope} is $\Newt(f) \coloneqq \conv(\{a_1,\dots,a_k\})$.
 Equivalently, $f$ is the \emph{support function} of $\Newt(f)$, that is, $f(x) = \max_{y \in \Newt(f)} y^\top x$.  
The map $\varphi \colon \mathcal{F}_d \to \mathcal{P}_d$, defined by $f \mapsto \Newt(f)$, is a bijection satisfying $\varphi(f+g) = \varphi(f) + \varphi(g)$ and $\varphi(\max\{f,g\}) = \conv(\varphi(f) \cup \varphi(g))$, where $+$ denotes pointwise addition or Minkowski sum, respectively.

\paragraph*{Parameterized Complexity.}
We assume basic knowledge on computational complexity theory.
Parameterized complexity is a multivariate approach to study the time complexity of computational problems \citep{Cygan15,DF13}.
A \emph{parameterized problem} $L\subseteq \Sigma^*\times \N$ consists of instances~$(x,k)$ where $x$ encodes a classical problem instance and $k$ is a \emph{parameter}.
A parameterized problem $L$ is \emph{fixed-parameter tractable} (contained in the class FPT) if it can be solved in $f(k)\cdot |x|^{\mathcal{O}(1)}$ time, where $f$ is an arbitrary function that only depends on $k$.
The class FPT is closed under so-called \emph{parameterized reductions}: A parameterized reduction from $L$ to $L'$ ($L\le_{\text{fpt}}L'$) is an algorithm that maps an instance $(x,k)$ in $f(k)\cdot |x|^{\mathcal{O}(1)}$ time to an instance $(x',k')$ such that $k'\leq g(k)$ for some function $g$ and $(x,k)\in L$ holds if and only if $(x',k')\in L'$.

The class XP contains all parameterized problems which can be solved in polynomial time for constant parameter values, that is, in $f(k)\cdot|x|^{g(k)}$ time, where $g$ is an arbitrary function that only depends on $k$.
It is known that FPT$\,\subsetneq\,$XP.

The \emph{W-hierarchy} consists of classes containing parameterized problems in XP which are presumably not fixed-parameter tractable. They are defined via the \textsc{Weighted Circuit Satisfiability} problem. A \emph{Boolean circuit} is a directed acyclic graph where each node of indegree~0 is an \emph{input node}, each indegree-1 node is a NOT-node, and each other node is an AND- or OR-node. Moreover, one node with outdegree~0 is labeled as \emph{output node}.
A node is \emph{large} if it has indegree at least 3.
The \emph{depth} of a circuit is the maximum length of a path from an input node to the output node.
The \emph{weft} of a circuit is the maximum number of large nodes on a path from an input node to the output node.
A Boolean circuit with~$n$ input nodes computes a Boolean function $f\colon\{0,1\}^n\to\{0,1\}$ in the natural way by assigning the values 0 or 1 to the input nodes. The central decision problem is the following. 

\problemdef{Weighted Circuit Satisfiability (WCS)}
{A Boolean circuit~$C$ and~$k\in\mathbb N$.}
{Does $C$ have a satisfying assignment of weight exactly~$k$ (that is, $k$ inputs are 1)?}

Let~$\mathcal C_{t,d}$ denote the class of Boolean circuits with weft~$t\ge 1$ and depth~$d\ge 1$ and let~WCS[$\mathcal C_{t,d}$] be the restriction of WCS to circuits in $\mathcal C_{t,d}$.

For $t\ge 1$, a parameterized problem~$L$ is contained in the class~W[$t$] if~$L\le_\text{fpt} \text{WCS}[\mathcal C_{t,d}]$ for some~$d\ge 1$.
A problem~$L$ is W[$t$]-hard if $Q\le_\text{fpt}L$ holds for every~$Q\in\,$W[$t$].
The class W[P] contains all parameterized problems~$L$ with $L\le_\text{fpt}\,$WCS.
The W-hierarchy satisfies the following inclusions
\[\text{FPT}\subseteq\text{W[1]}\subseteq\text{W[2]}\subseteq\cdots\subseteq\text{W[P]}\subseteq\text{XP}\] and it is widely believed that all inclusions are strict (at least one has to be).

\paragraph{Exponential Time Hypothesis.}
The Exponential Time Hypothesis (ETH) states that \mbox{3-SAT} on $n$ variables cannot be solved in $2^{o(n)}$ time~\citep{IP01}.
The ETH implies FPT$\,\neq\,$W[1] (which implies P$\,\neq\,$NP), as well as running time lower bounds: For example, \textsc{Clique} cannot be solved in $\rho(k)\cdot n^{o(k)}$ time, where $k$ is the size of the requested clique and $n$ is the number of nodes in the graph \citep{Cygan15}.

\subsection{Problem Definitions}

For given generator matrices $A\in\R^{d\times n}$ and $B\in\R^{d\times m}$\textcolor{black}{, and a scalar $L\in \R$}, we consider the following problems:
\begin{itemize}
    \item \textsc{Zonotope Containment}: Is $Z(A)\subseteq Z(B)$?
    \item \textsc{$L_p$-Max on Zonotopes}: Is $\max_{x\in Z(A)}\Vert x\Vert_p\geq L$?
\end{itemize}
For constant~$\ell\ge 1$, an $\ell$-layer ReLU network defined by weight matrices~$W_i\in\R^{n_i\times n_{i-1}}$ and biases~$b_i\in\R^{n_i}$ for $i\in[\ell]$, where $d\coloneqq n_0,\ldots,n_\ell\coloneqq 1\in\N^+$ that computes the function $f\colon\R^d\to \R,\;f(x)\coloneqq W_\ell\cdot (\phi_{W_{\ell-1},b_{\ell-1}}\circ \cdots \circ \phi_{W_1,b_1})(x)+b_\ell$, we consider the following problems:
\begin{itemize}
    \item \textsc{$\ell$-Layer ReLU Positivity}: Is there an~$x\in\R^d$ such that $f(x)>0$?
    \item \textsc{$\ell$-Layer ReLU Surjectivity}: Is $f$ surjective (that is, $\forall y\in\R\;\exists x\in\R^d : f(x)=y$)?
    \item \textsc{$\ell$-Layer ReLU $L_p$-Lipschitz Constant}: Is $L_p(f)\geq L$?
\end{itemize}
For the problems involving the $L_p$-Lipschitz constant, we assume access to an oracle that determines in polynomial time whether $\|v\|_p\geq L$ holds for $v\in \mathbb{R}^d,\;L\in \R$.

\section{Algorithmic Upper Bounds}
\label{sec:algorithms_and_containment}
All the problems above can easily be shown to be in XP for the parameter~$d$ (by simply enumerating all vertices of the zonotope or linear regions of the ReLU network).
In particular, the same problems with an unbounded number of layers (denoted by omitting the \textsc{$\ell$-Layer} prefix) are in~XP for the combined parameter $\ell d$.

\begin{theorem}\label{thm:enumeration_zonotopes}
\textsc{Zonotope Containment} and \textsc{$L_p$-Max on Zonotopes} can be solved in $n^{d-1}\cdot \poly(N)$ time (where $N$ is the input bit-length).
\end{theorem}

\begin{proof}
\textsc{Zonotope Containment}:
It is well-known that a zonotope $Z\subset \R^d$ with $n$ generators has $\mathcal{O}(n^{d-1})$ vertices which can be enumerated in $\mathcal{O}(n^{d-1})$ time~\citep{FFL05}.
Note that $Z$ is contained in another zonotope $Z'$ if and only if all vertices of $Z$ are contained in $Z'$.
Hence, by enumerating the vertices of $Z$ and checking containment in $Z'$ (e.g. by solving a linear program), we obtain an $\mathcal{O}(n^{d-1}\cdot \poly(N))$-time algorithm.

\textsc{$L_p$-Max on Zonotopes}:
We can enumerate all vertices of the zonotope in $n^{d-1}\cdot \poly(N)$ time and thus obtain an $n^{d-1}\cdot \poly(N)$ time algorithm using the $L_p$-norm oracle.
\end{proof}

\begin{theorem}\label{thm:enumeration_algorithms}
The following problems can be solved in $n^{(\ell - 1) d}\cdot \poly(N)$ time (where $n$ is the network width and $N$ is the input bit-length):
\begin{itemize}
    \item \textsc{$\ell$-Layer ReLU Positivity},
    \item \textsc{$\ell$-Layer ReLU Surjectivity},
    \item \textsc{$\ell$-Layer ReLU $L_p$-Lipschitz Constant},
    \item computing the maximum of an $\ell$-layer ReLU network over a polyhedron,
    \item deciding whether an $\ell$-layer ReLU network computes the zero function.
\end{itemize}
\end{theorem}

Before proving \Cref{thm:enumeration_algorithms}, we first prove that network verification for $\ell$-layer ReLU networks over polyhedra is solvable in $n^{(\ell - 1) d} \poly(N)$ time.
For this, we show that the set of full-dimensional activation regions can be enumerated in $n^{(\ell - 1) d} \poly(N)$ time with \Cref{alg:compute_regions}. 
The algorithm maintains a set~$\mathcal{R}_k$ whose elements encode the affine map and a halfspace description of every full-dimensional region on which the network truncated after the first $k$ hidden layers is affine.
In iteration $k+1$, each such region is partitioned by intersecting it with the hyperplane arrangement induced by the preactivations of the $(k+1)$-th hidden layer.

As regards notation, for a matrix $A\in\R^{n\times d}$ and a set $I\subseteq [n]$, $(A)_I$ denotes the submatrix consisting of the rows of $A$ whose row indices lie in $I$, and $[A]_I$ denotes the matrix $A$ where all rows whose row indices do not lie in $I$ are replaced with zero.

\begin{algorithm}
\caption{\textsc{ComputeRegions}}
\label{alg:compute_regions}
\begin{algorithmic}[1]
\Require{A ReLU network with weights $W_i\in \R^{n_i\times n_{i-1}}$ and biases $b_i\in \R^{n_i}$ for $i\in [\ell]$ computing $f:\R^d\to \R^m,\;f(x)=W_\ell \cdot (\phi_{W_{\ell-1},b_{\ell-1}} \circ \cdots \circ \phi_{W_{1},b_{1}})(x)+b_\ell$.}
\Ensure{The set $\mathcal{R}_{\ell-1}$ of $d$-dimensional activation regions.
Each region is encoded as a tuple $((A_r,a_r),(C_r,c_r))\in \mathcal{R}_{\ell-1}$ with $f(x)=A_rx+a_r$ for all $x\in \{z\in \R^d: C_rz+c_r\geq 0\}$.}
\State $\mathcal{R}_0\leftarrow \{((W_1,b_1),(0,0))\}$
\For{$k\in [\ell-1]$}
\State $\mathcal{R}_k\leftarrow \emptyset$
\For{$((A_r,a_r),(C_r,c_r))\in \mathcal{R}_{k-1}$}
\State $\mathcal{A}_r\leftarrow \textsc{CellPatterns}(A_r,a_r)$
\For{$I\in \mathcal{A}_r$}
\State $C\leftarrow (C_r, (A_r)_I,-(A_r)_{[n_k]\setminus I})^\top$, $c\leftarrow(c_r, (a_r)_I, -(a_r)_{[n_k]\setminus I})^\top$
\If{$\dim\{x\in \R^d: Cx+c\geq 0\}=d$}
\State $A\leftarrow W_{k+1} [A_r]_I$, $a\leftarrow W_{k+1}[a_r]_I+b_{k+1}$
\State $\mathcal{R}_k\leftarrow \mathcal{R}_k\cup \{((A,a),(C,c))\}$
\EndIf
\EndFor
\EndFor
\EndFor
\Return $\mathcal{R}_{\ell-1}$
\end{algorithmic}
\end{algorithm}
Given $A\in \R^{n\times d}$ and $a\in \R^n$, the function $\textsc{CellPatterns}(A,a)$ outputs a set $\mathcal{A}$ with one subset $I\subseteq [n]$ for each cell $C$ of the hyperplane arrangement $A_i x+a_i=0,\; i\in [n]$, where the corresponding cell is $C=\{x\in \R^d: A_{I} x +a_{I}\geq 0, A_{[n]\setminus I} x +a_{[n]\setminus I}\leq 0\}$.
\begin{proposition}
    \Cref{alg:compute_regions} is correct and runs in $n^{(\ell-1)d}\cdot \poly(N)$ time.
\end{proposition}
\begin{proof}
For correctness, we prove by induction that, for every $k\in \{0,1,\dots, \ell-1\}$, the set~$\mathcal{R}_k$ contains all full-dimensional activation regions after the first $k$ hidden layers.
Moreover, every element $((A_r,a_r),(C_r,c_r))\in \mathcal{R}_{k}$ satisfies $f_{k}(x)=A_rx+a_r$ for all $x\in P_r\coloneqq \{z\in \R^d: C_rz+c_r\geq 0\}$, where $f_k\colon\R^d\to \R^{n_k}$ denotes the function after $k$ hidden layers of the network, treating the $(k+1)$-th layer as the output layer.
This holds trivially for $k=0$.
Now, assume it holds for $k-1$.
Fix an $((A_r,a_r),(C_r,c_r))\in \mathcal{R}_{k-1}$ and an $I\subseteq [n_k]$.
The algorithm constructs $P_{r,I}\coloneqq\{x\in \R^d: Cx+c\geq 0\}$, which is the subset of $P_r$ on which the neurons of the $k$-th hidden layer with indices in $I$ are active (up to boundary points).
If $P_{r,I}$ is full-dimensional, then it is a full-dimensional activation region of $f_k$ and $f_k(x)=W_{k+1} [A_r]_I x + W_{k+1}[a_r]_I+b_{k+1}$ for all $x\in P_{r,I}$, which is stored in $\mathcal{R}_k$.
Conversely, every full-dimensional activation region of~$f_k$ is obtained by intersecting a region $r$ of $f_{k-1}$ with a cell of the hyperplane arrangement $(A_r)_jx+(a_r)_j=0,\; j\in [n_k]$.
Since $\textsc{CellPatterns}(A_r,a_r)$ enumerates all such cells, the set~$\mathcal{R}_k$ contains all full-dimensional activation regions of $f_k$, which proves the induction claim and therefore the correctness of \Cref{alg:compute_regions}.

For the running time, recall that an arrangement of at most $n$ hyperplanes in $\R^d$ has at most $\mathcal{O}(n^d)$ cells which can be enumerated in $\mathcal{O}(n^d)$ time \citep{edelsbrunner1986constructing}.
Hence each call to $\textsc{CellPatterns}$ runs in $n^d\cdot \poly(N)$ time.
For fixed $k\in [\ell-1]$, fixed $((A_r,a_r),(C_r,c_r))\in \mathcal{R}_{k}$, and fixed $I\in \mathcal{A}_r$, the algorithm performs a full-dimensionality test for a polyhedron in~$\R^d$ with at most $kn$ inequalities, which can be done in $\poly(N)$ time by linear programming \citep{Fukuda20}.
The matrix multiplications also take $\poly(N)$ time.
As $|\mathcal{R}_0|=1$ and $|\mathcal{R}_k|\leq |\mathcal{R}_{k-1}|\mathcal{O}(n^d)$ holds for all $k\in [\ell-1]$, we have $|\mathcal{R}_k|\leq \mathcal{O}(n^{kd})$ for all $k\in [\ell-1]$.
Therefore, the $k$-th iteration takes at most $\mathcal{O}(n^{(k-1)d})\cdot n^d \poly(N)=n^{kd}\poly(N)$ time.
Thus, the runtime of \Cref{alg:compute_regions} is $\sum_{k=1}^{\ell-1}n^{kd}\poly(N)=n^{(\ell-1)d}\poly(N)$.
\end{proof}

\Cref{alg:compute_regions} allows us to solve network verification for polyhedral constraints.

\begin{lemma}\label{lem:verification_for_polyhedra}
    Let $f\colon\R^d\to \R^m$ be an~$\ell$-layer ReLU network and let $P\subseteq \R^d$ and $Q\subseteq \R^m$ be polyhedra given in halfspace representation.
    Then, one can decide whether $f(P)\subseteq Q$ holds in $n^{(\ell - 1) d}\cdot \poly(N)$ time (where $n$ is the network width and $N$ the combined encoding size of the network and the polyhedra).
\end{lemma}
\begin{proof}
    Let $Q=\{x\in \R^m:v_i^\top x\leq u_i, i\in [k]\}$.
    Then, we have $f(P)\subseteq Q$ if and only if for every cell $C\in \Sigma_f$ and every constraint $i\in [k]$, we have
    \[
    u_i
    \geq \max\limits_{x\in C\cap P} v_i^\top f_C(x)
    =\max\limits_{x\in C\cap P} v_i^\top (A_C^\top x + b_C)
    =v_i^\top b_C + \max\limits_{x\in C\cap P} (A_C^\top v_i)^\top x,
    \]
    where $f_C$ is the affine linear function of $f$ restricted to $C$, that is, $f(x)=f_C(x)\coloneqq A_C^\top x+b_C$ for all $x\in C$.
    Note that the above condition can be verified by solving a linear program whose encoding size is polynomially bounded in $N$.
    Since linear programs can be solved in polynomial time and cells can be enumerated in $n^{(\ell - 1) d}\cdot \poly(N)$ time using \Cref{alg:compute_regions}, it follows that we can check whether $f(P)\subseteq Q$ holds in $n^{(\ell - 1) d}\cdot \poly(N)$ time.
\end{proof}

With \Cref{lem:verification_for_polyhedra} we can now prove \Cref{thm:enumeration_algorithms}.

\begin{proof}[Proof of \Cref{thm:enumeration_algorithms}]
Let $f\colon\R^d\to \R$ with $f(x)= W_\ell\cdot (\phi_{W_{\ell-1},b_{\ell-1}}\circ \cdots \circ \phi_{W_1,b_1})(x)+b_{\ell}$ be the function computed by an $\ell$-layer ReLU network and let $\Sigma_f$ be the corresponding polyhedral complex.
Further, let $n$ denote the width and $N$ the encoding size of the network and let $f_0\colon\R^d\to \R$ with $f_0(x)\coloneqq W_\ell\cdot (\phi_{W_{\ell-1},\mathbf{0}}\circ \cdots \circ \phi_{W_1,\mathbf{0}})(x)$ be the function of the ReLU network without biases.

\textsc{$\ell$-layer ReLU Positivity}:
Follows from applying \Cref{lem:verification_for_polyhedra} to $P=\R^d$ and $Q=(-\infty, 0]$, since the existence of a point $x\in \R^d$ with $f(x)>0$ is equivalent to $f(\R^d)\not \subseteq  (-\infty,0]$.

\textsc{$\ell$-layer ReLU Surjectivity}:
\citet[Lemma 14]{FGS25} show that for surjectivity, $f$ is surjective if and only if $f_0$ is surjective, which is equivalent to there being two points $r^+,r^-\in \R^d$ with $f_0(r^-)<0<f_0(r^+)$.
We can check this in $n^{(\ell - 1) d}\cdot \poly(N)$ time by applying the algorithm for \textsc{$\ell$-layer ReLU Positivity} to $f_0$ and $-f_0$.

\textsc{$\ell$-layer ReLU $L_p$-Lipschitz constant}:
Note that the $L_p$-Lipschitz constant of $f$ is equal to the maximum $L_q$-norm value (where $L_q$ is the dual norm of the $L_p$-norm, so $1/p+1/q=1$, or the $L_\infty$-norm for $p\in (0,1)$) of any gradient of the affine function that arises by restricting $f$ to a cell of $\Sigma_f$ (see \Cref{lem:lipschitz_region}).
Thus, by enumerating all cells of $\Sigma_f$, we obtain an $n^{(\ell - 1) d}\cdot \poly(N)$ time algorithm using the $L_q$-norm oracle.

\emph{Computing the maximum over a polyhedron $P$}:
Here, we assume that the maximum $\max_{x\in P}f(x)$ exists and $N$ denotes the combined encoding size of the network and the polyhedron $P$.
We can compute this maximum by enumerating cells of $\Sigma_f$ and solving the linear program $\max\limits_{x\in C\cap P}f_C(x)$ for each cell $C\in \Sigma_f$, where $f_C$ is the affine linear function of $f$ restricted to $C$, that is, $f(x)=f_C(x)\coloneqq a_C^\top x+b_C$ for all $x\in C$.
Note that the encoding size of the linear program is polynomially bounded in $N$.
Since linear programs can be solved in polynomial time and cells can be enumerated in $n^{(\ell - 1) d}\cdot \poly(N)$ time, it follows that $\max_{x\in P}f(x)=\max\limits_{C\text{ cell of }\Sigma_f}\max\limits_{x\in C\cap P}f_C(x)$ can be computed in $n^{(\ell - 1) d}\cdot \poly(N)$ time.

\emph{Checking for the zero function}:
Follows from \Cref{lem:verification_for_polyhedra} with $P=\R^d$ and $Q=\{0\}$.
\end{proof}

For the zonotope problems and the 2-layer ReLU network problems, we can even show containment in W[P].

\begin{theorem}\label{thm:WP-containment}
The following problems are contained in \textnormal{W[P]} when parameterized by~$d$:
\begin{itemize}
    \item \textsc{Zonotope Non-Containment}
    \item \textsc{$L_p$-Max on Zonotopes} for $p\in(0,\infty]$
    \item \textsc{$2$-Layer ReLU Positivity}
    \item \textsc{$2$-Layer ReLU Surjectivity},
    \item \textsc{$2$-Layer ReLU $L_p$-Lipschitz Constant} for $p\in(0,\infty]$.
\end{itemize}
\end{theorem}

\begin{proof}

In the following, let $N$ denote the number of input bits of the respective problem instance.
To show containment in W[P] with respect to $d$, it is sufficient to show that the problems can be solved by a nondeterministic algorithm that runs in~$\rho(d)\cdot N^{\mathcal{O}(1)}$ time and makes at most $\rho(d)\cdot\log(N)$ nondeterministic steps (for some function~$\rho$)~\citep{CFG05}.
In other words, it is enough to give a certificate for a yes-instance with $\rho(d)\cdot\log(N)$ bits which can be verified in $\rho(d)\cdot N^{\mathcal{O}(1)}$ time.
So, our general approach is to nondeterministically ``guess'' a cell of the associated polyhedral complex (which is a hyperplane arrangement) which serves as a certificate to be verified.

Hence, we first describe a certificate from which an inner point of a full-dimensional cell of a rational hyperplane arrangement can be computed.
Let
$
\mathcal H=\{H_1,\ldots,H_n\}
$
be a rational hyperplane arrangement in $\R^d$, where
$
H_i=\{x\in\R^d:h_i(x)=0\}
$
for affine linear functions $h_i\colon\R^d\to\R$.
For a sign pattern $\sigma\in\{\pm1\}^n$ of a cell, consider the linear program
\[
\max t \quad \text{s.t.}\quad \sigma_i h_i(x)\geq t \text{ for all } i\in[n],\quad 0\leq t\leq 1.
\]
If the cell is full-dimensional, then this linear program has an optimum which is strictly larger than $0$.
Hence, by the standard rational solution bound for linear programs \citep[Corollary~10.2a]{ilp_theory}, there is a rational upper bound with encoding size at most $B\coloneqq 4d^2(d+1)(N+1)$. Hence, every full-dimensional cell of $\mathcal H$ intersects $[-M,M]^d$ for $M\coloneqq 2^B$.
Let $C$ be the closure of a full-dimensional cell of $\mathcal H$.
Then
$
P_C\coloneqq  C\cap[-M,M]^d
$
is a full-dimensional polytope.
Hence, $P_C$ contains a full-dimensional simplex whose vertices are vertices of $P_C$ and whose barycenter lies inside $C$.
Each vertex of $P_C$ is determined by choosing at most $d$ hyperplanes among the hyperplanes $\mathcal H$ and the~$2d$ facets of the box $[-M,M]^d$.
Thus, a certificate for a full-dimensional cell consists of $d+1$ such vertices, each encoded by $d$ indices.
This requires $\mathcal{O}(d^2\log(n+2d))$ bits.

Given such a certificate, a verifier can compute the corresponding $d+1$ vertices, their barycenter~$x_0$, the active neurons on input $x_0$, and  the affine function that is computed on the cell containing~$x_0$.
All these computations take~$N^{\mathcal O(1)}$ time, since $M$ and all computed rational numbers have encoding length polynomially bounded in $N$.
Given the cell, the verifier proceeds as in the proof of \Cref{thm:enumeration_algorithms} to verify the respective ReLU problem instance in $N^{\mathcal O(1)}$ time.
This shows containment in W[P] for the stated ReLU problems.
Since \textsc{Zonotope Non-Containment} is equivalent to \textsc{$2$-Layer ReLU Positivity} without biases (see \Cref{sec:zonotope_containment_hardness}), the W[P]-containment of \textsc{$2$-Layer ReLU Positivity} also implies that \textsc{Zonotope Non-Containment} is contained in W[P].
Further, \textsc{$L_p$-Max on Zonotopes} is a special case of \textsc{$2$-Layer ReLU $L_p$-Lipschitz Constant} (see \citet{FGHS25}), hence it is clearly contained in W[P].
\end{proof}

When parameterized by $\ell$, we show that the ReLU network problems are in W[P] in the case where the input dimension $d=1$.
\begin{theorem}\label{thm:WP-containment_param_layers}
    \textsc{ReLU Positivity} and \textsc{ReLU $L_p$-Lipschitz Constant} for $p\in(0,\infty]$ with $d=1$ are contained in \textnormal{W[P]} when parameterized by~$\ell$, and \textsc{ReLU Surjectivity} with $d=1$ can be solved in polynomial time.
\end{theorem}
\begin{proof}
    As in the proof of \Cref{thm:WP-containment}, let $f\colon\R\to \R$ be the function computed by a ReLU network with input dimension one and let $N$ be the number of input bits.
    We show W[P]-containment with respect to $\ell$ by giving a nondeterministic algorithm that runs in time $\rho(\ell)\cdot N^{\mathcal{O}(1)}$ and makes at most $\rho(\ell)\cdot\log(N)$ nondeterministic steps, for some function~$\rho$.
    The main observation is that for $d=1$, every linear region is an interval, which can be encoded by choosing two neuron indices from each hidden layer, and, from this information, one can reconstruct the linear region and the affine function computed on it in polynomial time.
    
    We show that such a certificate exists by induction on the number of layers $\ell$.
    For $\ell=1$, the network has no hidden layer and the only linear region is $\mathbb{R}$; its affine function is given by $x\mapsto W_1x+b_1$.
    Now suppose that we have reconstructed an interval $I_{k-1}$ which is a linear region of the network truncated after layer $k-1$.
    Thus all functions computed by neurons in layer $k-1$ are affine functions of $x$ on $I_{k-1}$, and each preactivation of a neuron $i\in[n_k]$ in the $k$-th layer is affine on $I_{k-1}$, which we denote as $f_{k,i}(x)=\alpha_ix+\beta_i$.
    
    The linear regions of the network truncated after layer $k$ and contained in $I_{k-1}$ are formed by splitting $I_{k-1}$ at the points $-\beta_i/\alpha_i$ with $\alpha_i\neq 0$.
    Thus, each such region can be specified by its two endpoints, each of which is either an endpoint of $I_{k-1}$ or induced by a neuron in layer $k$.
    Given the two endpoints, the affine linear function of the interval can be computed in polynomial time by computing an interior point of the interval, e.g. the midpoint if both endpoints are finite, or a point a distance one away from the finite endpoint if the interval is unbounded.
    Evaluating the network at this point thus determines the active neurons and affine function on this interval.

    Overall, the certificate consists of two choices per hidden layer.
    In layer $k$, each choice is either one of $n_k$ neuron indices or one of two special symbols corresponding to the left and right endpoint of the interval of the previous layer.
    Thus, the number of possible indices is $\sum_{k=1}^{\ell-1}(n_k+2)\in \mathcal{O}(N)$ and encoding the certificate requires $2(\ell-1) \log(\sum_{k=1}^{\ell-1}(n_k+2))\in \mathcal{O}(\ell \log(N))$ bits.

    Given such a certificate, a verifier can compute in polynomial time the endpoints $\gamma_C^-$ and $\gamma_C^+$ and affine function $f_C(x)=\alpha_C x+\beta_C$ of the interval $C\subseteq \R$.
    Again, as in the proof of \Cref{thm:WP-containment}, a problem instance can be verified using~$C$ in $\poly(N)$ time (based on the algorithms from \Cref{thm:enumeration_algorithms}), which proves W[P]-containment.

    For \textsc{ReLU Surjectivity}, let $f_0$ denote the function computed by the corresponding network with all biases set to zero.
    By \citet{FGS25}, $f$ is surjective if and only if $f_0$ is surjective.
    Since $f_0$ is bias-free, it is positively homogeneous.
    Thus, for all $x\in \R_{\geq 0}$, we have $f_0(x)=x\cdot f(1)$ and $f_0(-x)=x\cdot  f(-1)$, which implies $f(\R)=\R_{\geq 0} \cdot f_0(1)\cup \R_{\geq 0} \cdot f_0(-1)$.
    Therefore, $f_0$ is surjective if and only if  $f_0(-1)$ and $f_0(1)$ (which can be computed in polynomial time) are non-zero and have opposite signs.
\end{proof}

\section{Reduction from \textsc{Multicolored Clique}}
\label{sec:gadget_construction}
In this section, we present a parameterized reduction which forms the basis for the hardness results for all our considered problems.
We reduce from the following problem.
\problemdef{Multicolored Clique}
{A graph $G=(V=V_1\dot\cup\cdots\dot\cup V_k, E)$, where each node in $V_i$ has color $i$.}
{Does $G$ have a $k$-colored clique (a clique with exactly one node of each color)?}
\textsc{Multicolored Clique} is NP-hard, W[1]-hard with respect to~$k$ and not solvable in $\rho(k)\cdot |V|^{o(k)}$ time for any computable function~$\rho$ assuming the ETH \citep{Cygan15}.

\begin{proposition}\label{lem:main_reduction}
    For every instance $(G=(V=V_1\dot\cup\cdots\dot\cup V_k, E),k)$ of \textsc{Multicolored Clique}, it is possible to construct in polynomial time a 2-layer ReLU network computing a function $f\colon\R^k\to \R$ such that $\max_{x\in \R^k}f(x)=k+\binom{k}{2}$ if and only if $G$ contains a $k$-colored clique and $\max_{x\in \R^k}f(x)\leq k+\binom{k}{2}- 1$ otherwise.
\end{proposition}

Before proving \Cref{lem:main_reduction}, we first introduce a definition and a useful lemma.

\begin{definition}
    A \emph{Sidon set} is a set $A=\{a_1,\dots, a_m\}\subseteq \mathbb Z^+$ where all pairwise sums $a_i+a_j$ with $i\leq j$ are different.
\end{definition}
For a survey on Sidon sets, we refer to~\citep{o2004complete}.
The greedy Sidon set, introduced by \citet{mian1944b2}, is recursively constructed as follows: Let $a_1\coloneqq 1$, and for $n>1$, let $a_n$ be the smallest nonnegative integer such that $\{a_1,\dots, a_n\}$ is a Sidon set (see \href{https://oeis.org/A005282}{A005282}).
\citet{stohr1955geloste} noted that $a_n\in \mathcal{O}(n^3)$.
Moreover, note that the greedy Sidon set of size $n$ can be computed in $n^{\mathcal{O}(1)}$ time.
We use the following elementary lemma.
\begin{lemma}\label{lem:sidon_set_uniqueness}
    Let $W_1,\dots, W_k$ be a partition of a Sidon set $A$ into disjoint subsets.
    Then, for every pair $(i,j)\in \binom{[k]}{2}$, all pairwise sums $a+b$ with $a\in W_i$ and $b\in W_j$ are different.
\end{lemma}

\begin{proof}[Proof of \Cref{lem:main_reduction}.]
Let $(G=(V=V_1\dot\cup\ldots\dot\cup V_k, E),k)$ be an instance of \textsc{Multicolored Clique}, where $V_c=\{v_{c,1},\dots, v_{c,n_c}\}$ for $c\in[k]$ and $E=\bigcup_{(r,l)\in \binom{[k]}{2}}E_{r,l}$, where $E_{r,l}$ denotes the set of edges whose nodes have color $r$ and $l$.
Further, let $A$ be the greedy Sidon set of size $|V|$ and let $W_1,\dots, W_k$ be a partition of $A$ into $k$ disjoint subsets such that $|W_i|=n_i$ holds.
Note that this allows us to assign each node $v_{c,i}$ to a unique element $\omega_{c,i}$ of $A$, namely the $i$-th element of $W_c$.
For every edge $\{v_{r,i}, v_{l,j}\}$, we define the constant $\omega_{r,i,l,j}:=\omega_{r,i}+\omega_{l,j}$.
\Cref{lem:sidon_set_uniqueness} implies that no two edges get the same constant: the value $\omega_{r,i,l,j}$ uniquely determines the edge $\{v_{r,i}, v_{l,j}\}$.
We construct a ReLU network with $k$ input variables $x_1,\dots, x_k$ and $3(|V|+|E|)$ hidden neurons as follows.

For every color $c\in[k]$, we introduce a \emph{node selection} gadget which ensures that the input value~$x_c$ encodes a node in~$V_c$.
To this end, we create a ``penalty function''
$p_c\colon\R\to [0,1]$  (see \Cref{fig:penalty_function}) that has $n_c$ narrow spikes around the value $\omega_{c,i}$ (that is, it goes up from 0 to 1 and down to 0 again) for each $v_{c,i}\in V_c$ and is zero everywhere else:
\begin{equation*}
    p_c(x)\coloneqq\begin{cases}
        8(x-\omega_{c,i}+\frac{1}{8}), & \text{ if }x\in [\omega_{c,i}-\frac{1}{8}, \omega_{c,i}], \quad i\in [n_c]\\
        1-8(x-\omega_{c,i}), & \text{ if }x\in (\omega_{c,i},\omega_{c,i}+\frac{1}{8}], \quad i\in [n_c]\\
        0, & \text{ if }x\notin \bigcup_{i\in [n_c]}[\omega_{c,i}-\frac{1}{8},\omega_{c,i}+\frac{1}{8}]
    \end{cases}
\end{equation*}
The penalty function $p_c$ can be implemented with $3n_c$ hidden neurons:
\begin{equation*}
p_c(x)=\sum_{i\in [n_c]}(\max(0, 8(x-\omega_{c,i}+\frac{1}{8}))
-\max(0, 16(x-\omega_{c,i}))
+\max(0, 8(x-\omega_{c,i}-\frac{1}{8}))).
\end{equation*}
\begin{figure}
  \centering
  \begin{subfigure}[t]{0.55\textwidth}
    \centering
    \begin{minipage}[c][\contentH][c]{\linewidth}
      \resizebox{\linewidth}{!}{\usebox{\leftbox}}
    \end{minipage}
    \phantomsubcaption\label{fig:spike}
    \subcaption*{Figure \thefigure:
    Spike function $s_{r,l}$ for a colored graph (top left).
    Node labels: $\omega_{r,1}=1,\omega_{r,2}=2,\omega_{l,1}=4$, $\omega_{l,2}=8$.
    Edge labels: $\omega_{r,1,l,1}=5, \omega_{r,2,l,1}=6, \omega_{r,1,l,2}=9$,  $\omega_{r,2,l,2}=10$.
    }
  \refstepcounter{figure}
  \end{subfigure}
  \hfill
  \begin{subfigure}[t]{0.44\textwidth}
    \centering
    \begin{minipage}[c][\contentH][c]{\linewidth}
      \resizebox{\linewidth}{!}{\usebox{\rightbox}}
    \end{minipage}    \phantomsubcaption\label{fig:penalty_function}
    \subcaption*{Figure \thefigure: Penalty function $p_c$.}
  \end{subfigure}
\end{figure}
Next, we introduce an \emph{edge verification} gadget which verifies that each pair of nodes selected by the node selection gadgets is connected by an edge. For every pair of colors $(r,l)\in \binom{[k]}{2}$, we define a ``spike function'' $s_{r,l}\colon\R^2\to [0,1]$
(see \Cref{fig:spike}) that is zero everywhere except for a set of $|E_{r,l}|$ parallel stripes in which $s_{r,l}$ forms a spike.
\begin{equation*}
    s_{r,l}(x,y)\coloneqq\begin{cases}
        4(x+y-\omega_{r,i,l,j}-\frac{1}{4}), & \text{ if }x+y\in [\omega_{r,i,l,j}-\frac{1}{4}, \omega_{r,i,l,j}], \quad \{v_{r,i}, v_{l,j}\}\in E_{r,l}\\
        1-4(x+y-\omega_{r,i,l,j}), & \text{ if }x+y\in (\omega_{r,i,l,j},\omega_{r,i,l,j}+\frac{1}{4}], \quad \{v_{r,i}, v_{l,j}\}\in E_{r,l}\\
        0, & \text{ if }x+y\notin \bigcup_{\{v_{r,i}, v_{l,j}\}\in E_{r,l}}[\omega_{r,i,l,j}-\frac{1}{4},\omega_{r,i,l,j}+\frac{1}{4}]
    \end{cases}
\end{equation*}
Note that \Cref{lem:sidon_set_uniqueness} guarantees that $s_{r,l}$ is well-defined, as the sums $\omega_{r,i}+\omega_{l,j}=\omega_{r,i,l,j}$ for $\{v_{r,i}, v_{l,j}\}\in E_{r,l}$ are all integral and different, so no two intervals $[\omega_{r,i,l,j}-\frac{1}{4},\omega_{r,i,l,j}+\frac{1}{4}]$ and $[\omega_{r',i',l',j'}-\frac{1}{4},\omega_{r',i',l',j'}+\frac{1}{4}]$ overlap for different $(r,i,l,j)$ and $(r',i',l',j')$.
Thus, the spike function attains value $1$ if and only if the sum of its inputs is equal to the weight of some edge.
The spike function can be implemented with $3|E_{r,l}|$ hidden neurons:
\begin{align*}
s_{r,l}(x,y)=
\sum_{\{v_{r,i}, v_{l,j}\}\in E_{r,l}}(\max(0, 4(x+y-\omega_{r,i,l,j}+\frac{1}{4}))
-&\max(0, 8(x+y-\omega_{r,i,l,j}))\\
+&\max(0, 4(x+y-\omega_{r,i,l,j}-\frac{1}{4}))).
\end{align*}
By computing all penalty and spike functions in parallel and summing them up at the output neuron, we obtain a ReLU network that computes the function $f\colon\R^k\to \R$ with
\begin{equation*}
    f(x_1,\dots,x_k)=\sum_{(r,l)\in \binom{[k]}{2}}s_{r,l}(x_r,x_l)+\sum_{c\in [k]}p_c(x_c).
\end{equation*}
\Cref{fig:color_pair_functions} illustrates how the sum of the spike function $s_{r,l}$ and penalty functions $p_r,p_l$ behaves for a fixed color pair.
\begin{figure}[t]
\centering
\includegraphics[width=0.65\textwidth, page=4]{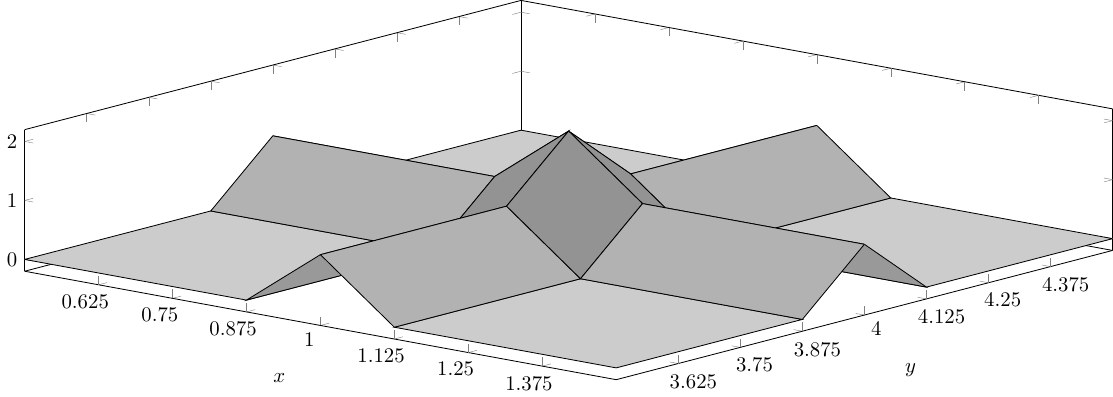}
\includegraphics[width=0.275\textwidth, page=5]{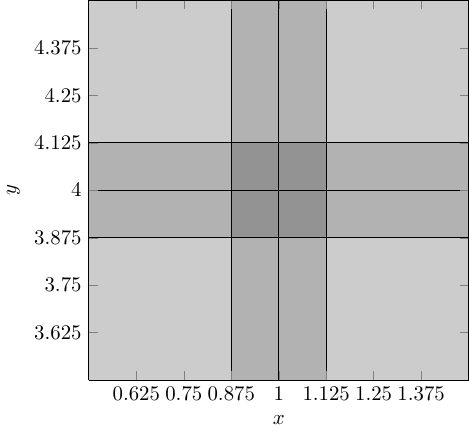}
\includegraphics[width=0.65\textwidth, page=6]{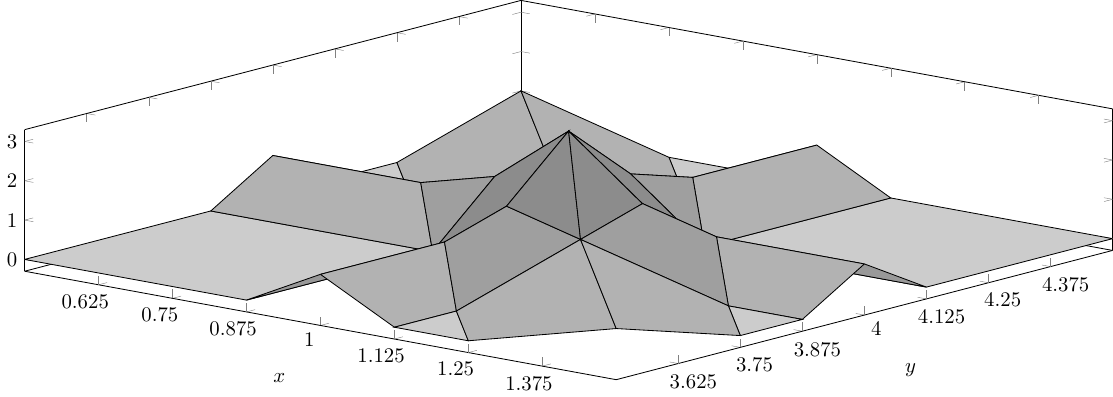}
\includegraphics[width=0.275\textwidth, page=7]{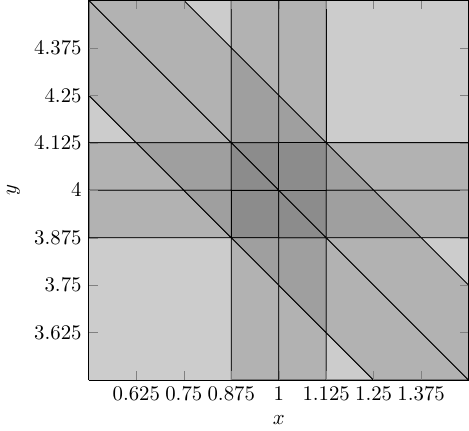}
\caption{
Illustration of the area around node values for a fixed color pair for non-adjacent and adjacent nodes.
The values are based on the example given in \Cref{fig:spike}.
For non-adjacent nodes, only the two penalty functions intersect, while for adjacent nodes, the penalty functions intersect also with the spike function.
}
\label{fig:color_pair_functions}
\end{figure}
Since every spike and penalty function is lower bounded by 0 and upper bounded by 1, it follows that $f$ is lower bounded by 0 and upper bounded by $k+\binom{k}{2}$.

First, we show that the existence of a $k$-colored clique in $G$ implies $\max_{x\in \R^k}f(x)=k+\binom{k}{2}$.
Suppose that $\{v_{1,a_1},\dots, v_{k,a_k}\}\subset V$ forms a $k$-colored clique in $G$.
Then, we claim that the point $x^*=(\omega_{1,a_1},\dots, \omega_{k,a_k})$ is a point with $f(x^*)=k+\binom{k}{2}$.
First, note that $p_c(x^*_c)=1$ holds for all $c\in [k]$.
Further, $s_{r,l}(x^*_r,x^*_l)=1$ holds for each pair of colors $(r,l)\in \binom{[k]}{2}$, since $\{v_{r, a_r}, v_{l,a_l}\}$ is an edge in $E_{r,l}$.
Thus, we have $f(x^*)=k+\binom{k}{2}$.

Now, we show that if there is a point $x^*\in \R^k$ with $f(x^*)>k+\binom{k}{2}-1$, then $G$ has a $k$-colored clique.
Suppose that $x^*\in \R^k$ is such a point.
Therefore, we must have $p_c(x^*_c)>0$ for every $c\in [k]$ and $s_{r,l}(x^*_r,x^*_l)>0$ for every pair $(r,l)\in \binom{[k]}{2}$, since otherwise $f(x^*)\leq k+\binom{k}{2}-1$ holds.
By definition, $p_c(x^*_c)>0$ implies that there is exactly one element $v_{c,a_c}\in V_c$ with $x^*_c\in (\omega_{c,a_c}-\frac{1}{8},\omega_{c,a_c}+\frac{1}{8})$.
In other words, the input $x^*_c$ corresponds to the node $v_{c,a_c}$ for every $c\in [k]$.
We now claim that $\{v_{1,a_1},\dots, v_{k,a_k}\}$ forms a $k$-colored clique in $G$.
To see this, fix an arbitrary pair $(r,l)\in \binom{[k]}{2}$.
First, note that $x^*_r\in (\omega_{r,a_r}-\frac{1}{8},\omega_{r,a_r}+\frac{1}{8})$ and $x^*_l\in (\omega_{l,a_l}-\frac{1}{8},\omega_{l,a_l}+\frac{1}{8})$ implies $x^*_r+x^*_l\in (\omega_{r,a_r,l,a_l}-\frac{1}{4},\omega_{r,a_r,l,a_l}+\frac{1}{4})$.
Recall that for inputs $(x,y)$ with $x+y\in(\omega_{r,a_r,l,a_l}-\frac{1}{4},\omega_{r,a_r,l,a_l}+\frac{1}{4})$, the spike function $s_{r,l}$ has by definition a strictly positive output $s_{r,l}(x,y)>0$ if $\{v_{r,a_r},v_{l,a_l}\}$ is an edge in $E_{r,l}$, and output zero $s_{r,l}(x,y)=0$ otherwise.
Since $s_{r,l}(x^*_r,x^*_l)>0$ holds by assumption, this means that $\{v_{r,a_r},v_{l,a_l}\}$ is an edge in $E_{r,l}$.
As $(r,l)$ was an arbitrary pair, all pairs of nodes in $\{v_{1,a_1},\dots, v_{k,a_k}\}$ are adjacent, which shows that the set is indeed a $k$-colored clique.
\end{proof}

In the following, we will use modifications of this construction to prove our hardness results.
All our (parameterized) reductions are in fact \emph{polynomial-time reductions} and thus also prove NP-hardness.
We will only state this explicitly if the NP-hardness of the problem was not previously known.

\section{Hardness of Network Verification Problems}
\label{sec:verification}

We first prove W[1]-hardness of \textsc{2-Layer ReLU Positivity} parameterized by~$d$ (NP-hardness of \textsc{2-layer ReLU Positivity} was established by \citet{FGS25}) and later extend it to W[$\ell$]-hardness for \textsc{$(\ell+1)$-Layer ReLU Positivity}.
For W[1]-hardness, we use the reduction from \Cref{lem:main_reduction}, while for W[$\ell$]-hardness, we use another reduction.
However, both reductions rely on the existence of biases.
To extend the hardness result to other problems, we need to show the stronger statement that \textsc{$(\ell+1)$-layer ReLU Positivity} remains W[$\ell$]-hard even when all biases are zero.

To motivate the construction, first consider a ReLU network with biases computing $f:\R^d\to \R$.
A simple way to remove biases is to add one extra input variable~$y$ and replace each bias $b$ with $b y$.
To replace biases $b$ with $by$ that occur in deeper layers, we can introduce two extra ReLU neurons in each hidden layer to propagate the identity $y=\max(0,y)-\max(0,-y)$ to each layer.
This construction gives a bias-free ReLU network computing $h:\R^{d+1}\to \R$ with $h(x,1)=f(x)$ for all $x\in \R^d$.
It is well known that bias-free ReLU networks compute positively homogeneous functions, that is, $h(\lambda x,\lambda y)=\lambda h(x,y)$ holds for all $\lambda\geq 0$.
Thus, if $f(x^*)>0$, then also $h(x^*,1)>0$.
Conversely, if $f(x)\leq 0$ for all $x\in \R^d$, then for every $(x,y)\in \R^d\times \R_{>0}$, we have $h(x,y)=yf(x/y)\leq 0$.
However, this implication does not hold for inputs $(x,y)\in \R^d\times \R_{<0}$.

One way to restore symmetry between $y>0$ and $y<0$ is to replace every bias $b$ with $b|y|$.
Then the resulting function $h_{\rm abs}$ satisfies $h_{\rm abs}(x,1)=f(x)=h_{\rm abs}(x,-1)$ for all $x\in \R^d$.
The drawback is that computing $|y|=\max(0,y)+\max(0,-y)$ requires an additional hidden layer before $|y|$ can be used.

We therefore use a ``mixed'' construction that avoids an extra hidden layer and restores the symmetry between $y>0$ and $y<0$.
\begin{definition}\label{def:homogenization}
    Let $f\colon\R^d\to\R$ be computed by a ReLU network $\mathcal{N}$.
    The \emph{homogenization} of $\mathcal{N}$ is obtained by adding an extra input variable~$y$ to $\mathcal{N}$ and modifying the biases as follows: biases $b$ in the first hidden layer are replaced by $b\cdot y$, while biases $b$ in all subsequent hidden layers and in the output layer are replaced by $b|y|=b\max(0,y)+b\max(0,-y)$ using two extra ReLU neurons in each hidden layer that compute $\max(0,y)$ and $\max(0,-y)$.
\end{definition}

\begin{figure}
\centering
\includegraphics[width=0.26\textwidth, page=8]{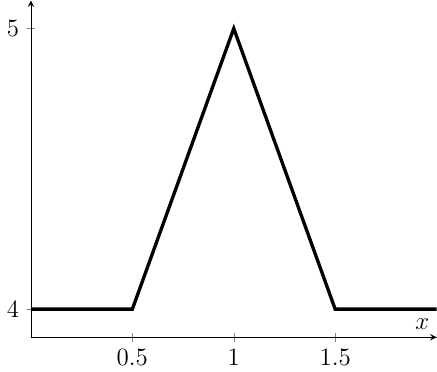}
\hfill
\includegraphics[width=0.26\textwidth, page=9]{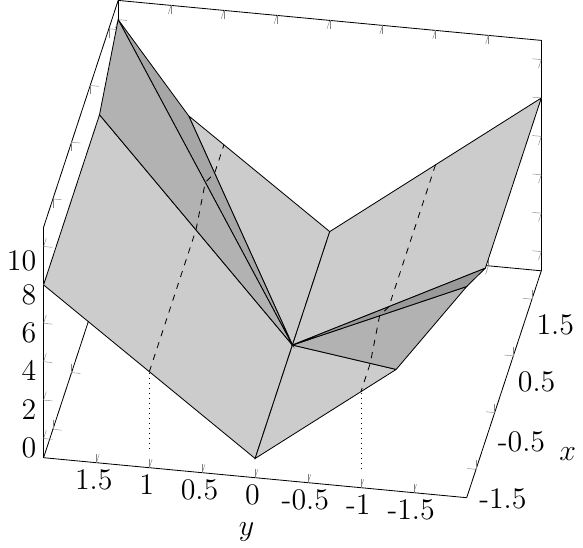}
\hfill
\includegraphics[width=0.25\textwidth, page=10]{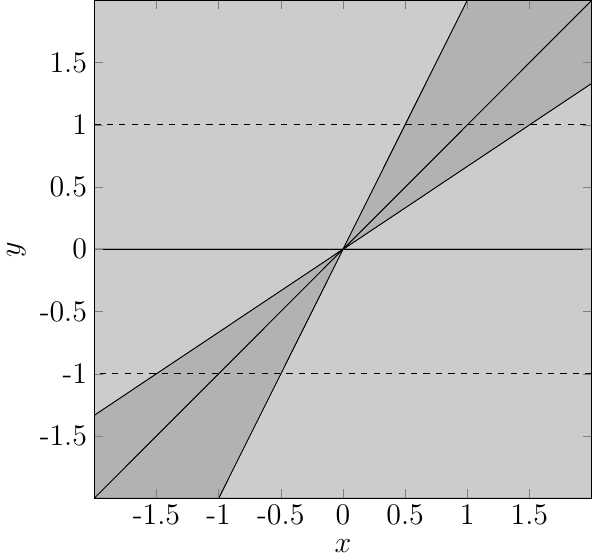}
\caption{
Homogenization: the function $\max(0,2x-1)-\max(0,4x-4)+\max(0,2x-3)+4$ (left) is turned into $\max(0,2x-y)-\max(0,4x-4y)+\max(0,2x-3y)+4|y|$ (right).
}
\label{fig:homogenization}
\end{figure}

\Cref{fig:homogenization} illustrates the geometric effect of homogenization on the function of a 2-layer ReLU network.
Before we start with the W[1]-hardness proof, we first prove an auxiliary lemma that allows us to remove biases from our constructions.

\begin{lemma}\label{lem:homogenization_function}
    Fix $\ell\geq 2$.
    Let $f\colon\R^d\to\R$ with $f(x)= W_\ell\cdot (\phi_{W_{\ell-1},b_{\ell-1}}\circ \cdots \circ \phi_{W_1,b_1})(x)+b_{\ell}$ be the function computed by an $\ell$-layer ReLU network, and let $h\colon\R^{d+1}\to \R$ be the function computed by its homogenization.
    If $x\in \R^d$ satisfies $W_2(W_1x+b_1)=0$, then
    \[h(x,1)=h(-x,-1).\]
\end{lemma}
\begin{proof}
    Let $x\in \R^d$ satisfy $W_2(W_1x+b_1)=0$ and let $h_l(x,y)$ denote the pre-activation function of the $l$-th layer of the homogenized network, e.g., the pre-activation in the second layer is
    \[
    h_2(x,y)=W_2\max\{0,W_1x+b_1y\}+b_2|y|.
    \]
    We have
    \begin{align*}
    h_2(x,1)-h_2(-x,-1)=W_2(\max\{0,W_1x+b_1\}-\max\{0,-(W_1x+b_1)\})=W_2(W_1x+b_1)=0,
    \end{align*}
    which gives $h_2(x,1)=h_2(-x,-1)$.
    This equality propagates through the remaining layers.
    Suppose that for some $l\in \{3,\dots,\ell-1\}$, we have $h_{l-1}(x,1)=h_{l-1}(-x,-1)$.
    Then, we have
    \begin{align*}
    h_l(x,1)
    =W_l\max\{0,h_{l-1}(x,1)\}+b_l=
    W_l\max\{0,h_{l-1}(-x,-1)\}+b_l
    =h_l(-x,-1).
    \end{align*}
    By induction, we have $h_{\ell-1}(x,1)=h_{\ell-1}(-x,-1)$ and thus
    \[
    h(x,1)=W_\ell\max\{0,h_{\ell-1}(x,1)\}+b_\ell = W_\ell\max\{0,h_{\ell-1}(-x,-1)\}+b_\ell=h(-x,-1).
    \]
\end{proof}

We are now ready to prove the following theorem.

\begin{theorem}\label{thm:positivity_hardness_without_bias}
    \textsc{2-layer ReLU Positivity} is \textnormal{W[1]}-hard with respect to $d$ and not solvable in $\rho(d)\cdot N^{o(d)}$ time (where $N$ is the input bit-length) for any function $\rho$ assuming the ETH, even if all biases are zero.
\end{theorem}

\begin{proof}
    We give a parameterized reduction from \textsc{Multicolored Clique}.
    Let $(G=(V=V_1\dot\cup\ldots\dot\cup V_k, E),k)$ be an instance of \textsc{Multicolored Clique}, and let $f\colon\R^k\to \R$ be the function of the network constructed in the proof of \Cref{lem:main_reduction}.
    Next, we modify the network by setting the bias of the output node to $1-k-\binom{k}{2}$.
    Let $g\colon\R^k\to \R$ be the function of this modified network and let $h\colon\R^{k+1}\to \R$ be the function computed by the homogenization of this modified network.
    Hence, we have $h(x,1)= g(x)= f(x)+1-k-\binom{k}{2}$ for every $x\in \R^k$.
    Note that $h(-x,-1)=h(x,1)$ holds for every $x\in \R^k$, which follows from \Cref{lem:homogenization_function}, as $g$ satisfies the assumption of \Cref{lem:homogenization_function} for all $x\in \R^k$.
    Since the ReLU network corresponding to $h$ has no biases, $h$ is positively homogeneous.

    By \Cref{lem:main_reduction}, $G$ has a $k$-colored clique if and only if $g$ has a \emph{positive point}, since $\max_{x\in \R^k}g(x)=1$ holds if $G$ has a $k$-colored clique and $\max_{x\in \R^k}g(x)\leq 0$ otherwise.
    We claim that $h$ has a positive point if and only if $g$ has a positive point.
    To see this, first observe that if $g(x^*)>0$, then $h(x^*,1)>0$.
    Conversely, if $h(x^+,y^+)>0$, then by positive homogeneity we have $0<\frac{1}{|y^+|}h(x^+,y^+)=h(\frac{x^+}{|y^+|},\text{sgn}(y^+))=g(\text{sgn}(y^+)\cdot \frac{x^+}{|y^+|})$.
    Note that $y^+= 0$ is not possible, since $h(x,0)=0$ for every $x\in \R^k$ by construction, as deleting biases leads to the cancellation of all terms in the spike and penalty functions.
\end{proof}

Next, we generalize \Cref{thm:positivity_hardness_without_bias} to W[$\ell$]-hardness for~$\ell+1$ layers.

\begin{theorem}\label{thm:Wl-hardness}
For every fixed~$\ell\ge 1$, \textsc{$(\ell+1)$-Layer ReLU Positivity} is \textnormal{W[$\ell$]}-hard with respect to~$d$  and not solvable in $\rho(d)\cdot N^{o(d)}$ time (where $N$ is the input bit-length) for any function $\rho$ assuming the ETH, even if all biases are zero.
\end{theorem}

The proof is based on parameterized reductions from
the \textsc{Weighted (Anti)Monotone $t$-Normalized SAT} problem.
For~$t\ge 1$, a Boolean formula is \emph{$t$-normalized} if it is a conjunction of disjunctions of conjunctions of ... of literals, alternating $t-1$ times between conjunctions and disjunctions.
    A formula is \emph{monotone} (\emph{antimonotone}) if all literals are positive (negative).

    \problemdef{Weighted (Anti)Monotone $t$-Normalized SAT}
{A (anti)monotone $t$-normalized Boolean formula~$F$ and~$k\in\N$.}
{Does $F$ have a satisfying assignment of weight~$k$?}

Notably, the proof of W[$\ell$]-hardness of \textsc{$(\ell+1)$-Layer ReLU Positivity} for $\ell\geq 2$ (two or more hidden layers) is conceptually much simpler than the proof of W[1]-hardness of \textsc{$2$-Layer ReLU Positivity} (one hidden layer): for $\ell\geq 2$, we can use the fact that ReLU networks of sufficient depth can simulate Boolean formulas.

\begin{proof}[Proof of \Cref{thm:Wl-hardness}]
    The case~$\ell=1$ is \Cref{thm:positivity_hardness_without_bias}.
    Moreover, note that concatenating this construction with layers computing the identity map yields the ETH-based runtime lower bound for \textsc{$(\ell+1)$-Layer ReLU Positivity} for every fixed $\ell\geq 1$.
    
    \textbf{Case I: Even $\ell \ge 2$.} We give a polynomial-time parameterized reduction from the~W[$\ell$]-hard \textsc{Weighted Monotone $\ell$-Normalized SAT} problem \citep{DF13}.
    
    Given an instance~$(F,k)$ with a monotone $\ell$-normalized formula~$F$
    with variables $z_1,\ldots,z_n$, we construct a ReLU network with~$\ell+1$ layers and input dimension~$k$. The idea is that the~$k$ inputs specify the indices of the variables which are set to true in a weight-$k$ satisfying assignment for~$F$. The first hidden layer decodes the $k$ input numbers into binary values for the~$n$ variables and the subsequent~$\ell$ layers compute the disjunctions and conjunctions.

 To this end, the first hidden layer contains three neurons for every $c\in [k]$ and~$i\in[n]$ computing the values $\max(0, 8(x_c-i+\frac{1}{8}))$, $\max(0, 16(x_c-i))$ and $\max(0, 8(x_c-i-\frac{1}{8}))$, which can be combined in the second layer to a function $p_{c,i}\colon\R^k\to[0,1]$ defined as
    \begin{equation*}
p_{c,i}(x)\coloneqq\max(0, 8(x_c-i+\frac{1}{8}))
-\max(0, 16(x_c-i)) +\max(0, 8(x_c-i-\frac{1}{8})),
\end{equation*}
    which is positive if and only if $x_c\in(i-1/8,i+1/8)$ (see \Cref{fig:pci_function_plot}).
\begin{figure}
    \centering
    \includegraphics[width=0.325\textwidth, page=11]{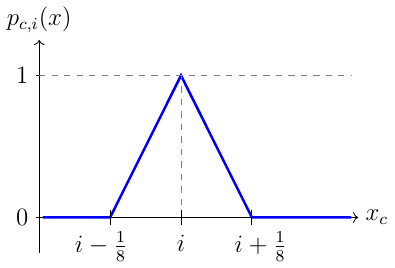}
    \hfill
    \includegraphics[width=0.31\textwidth, page=12]{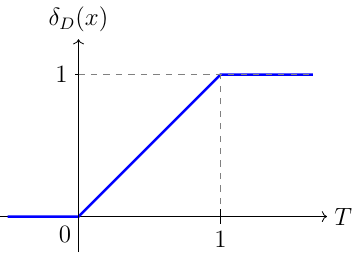}
    \hfill
    \includegraphics[width=0.325\textwidth, page=13]{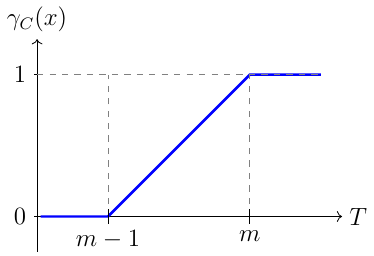}
    \caption{Plot of $p_{c,i}$ (left), $\delta_D$ for $T=\sum_{j=1}^m \gamma_{C_j}(x)$ (center), $\gamma_C$ for $T=\sum_{j=1}^m\delta_{D_j}(x)$ (right).}
    \label{fig:pci_function_plot}
\end{figure}

The second hidden layer then ``simulates'' the innermost disjunctions of~$F$ over positive literals using the values of the $p_{c,i}$ functions as follows: To compute the value of a disjunction $D=z_{i_1}\vee\cdots\vee z_{i_m}$, we use two neurons whose output values are combined (in the third layer) to the function 
\[
\delta_D(x)
\coloneqq\max\left(0,\sum_{j=1}^m \sum_{c\in[k]}p_{c,i_j}(x)\right) - \max\left(0,\sum_{j=1}^m \sum_{c\in[k]}p_{c,i_j}(x)-1\right)\in[0,1],
\]
which attains a positive value if and only if at least one~$p_{c,i_j}(x)$ is positive (see \Cref{fig:pci_function_plot}).

    The next layer simulates conjunctions  $C=D_1\wedge\cdots\wedge D_m$ using the disjunction values~$\delta_{D_1},\ldots,\delta_{D_m}$ from the previous layer as follows:
    \[\gamma_C(x)\coloneq\max(0,\sum_{j=1}^m\delta_{D_j}(x)-m+1)-\max(0,\sum_{j=1}^m\delta_{D_j}(x)-m)\in[0,1].\]
    Note that $\gamma_C(x)>0$ only if $\delta_{D_j}(x)>0$ for all $j\in[m]$ since $\delta_{D_j}(x)\le 1$ for all $x\in \R^k$ (see \Cref{fig:pci_function_plot}).

    Again, the next layer simulates disjunctions $D=C_1\lor \cdots \lor C_m$ using the construction of the second hidden layer (simply replace $\sum_{c\in [k]} p_{c,i_j}(x)$ with $\gamma_{C_j}(x)$ in the formula)
    \[
    \delta_D(x):=\max\left(0,\sum_{j=1}^m \gamma_{C_j}(x)\right)-\max\left(0,\sum_{j=1}^m \gamma_{C_j}(x) -1\right)\in [0,1],
    \]
    which again attains a positive value if and only if at least one $\gamma_{C_j}(x)$ is positive (see \Cref{fig:pci_function_plot}).
    
    Now, we continue the construction by repeatedly adding layers computing disjunctions and conjunctions according to the formula~$F$ (using the conjunction and disjunction values from the previous layer).
    Let the outermost conjunction of~$F$ be $D_1\wedge\cdots\wedge D_m$. The output layer computes the final value
    \[f(x)\coloneqq\sum_{j=1}^m\delta_{D_j}(x)-m+1.\]

    The correctness is straightforward: If $(F,k)$ is a yes-instance, then let $i_1,\ldots,i_k$ be the indices of the variables which are set to true in a satisfying assignment~$\alpha$ for~$F$. We claim that the input $x^*\coloneqq(i_1,\ldots,i_k)$ yields $f(x^*)>0$. To see this, note that $p_{j,i_j}(x^*)=1$ for every~$j\in[k]$. Hence, for the second layer, $\delta_D(x^*)=1$ holds for each innermost disjunction~$D$ of~$F$ which evaluates to true under $\alpha$. Clearly, in the third layer then also $\gamma_C(x^*)=1$ holds for all following conjunctions~$C$ in~$F$ which are satisfied by~$\alpha$. Thus, repeating this argument yields $f(x^*)=1$ since~$F$ is satisfied by~$\alpha$.
    
    Now, assume that $f(x^*)>0$ holds for some~$x^*\in\R^k$. Let~$I\coloneqq\{i\in[n]\mid \exists \,c\in[k]:p_{c,i}(x^*)>0\}$.
    Note that~$|I|\le k$ since $p_{c,i}(x^*)>0$ implies~$p_{c,i'}(x^*)\le 0$ for all other~$i'\in[n]$.
    We claim that setting the variables~$z_i$ with~$i\in I$ to true yields a satisfying assignment for~$F$. (Note that, if~$|I|<k$, then we can simply set~$k-|I|$ arbitrary other variables to true to obtain a weight-$k$ satisfying assignment since~$F$ is monotone.)
    To verify the claim, note that $f(x^*)>0$ implies $\sum_{j=1}^m\delta_{D_j}(x^*) > m-1$ which implies $\delta_{D_j}(x^*)>0$ for all $j\in[m]$ since $\delta_{D_j}(x)\le 1$ for all~$x\in\R^k$. This implies for each disjunction~$D_j$ that at least one of its corresponding input values from the previous layer (that is, a value of a~$\gamma_C$ function) is positive. This again implies that all the corresponding input values of this function from the layer before are positive. Hence, it follows inductively that~$I$ yields a satisfying assignment.

    \textbf{Removing biases in Case I.}
    Let $h\colon\R^{k+1}\to \R$ be the homogenization (see \Cref{def:homogenization}) of the constructed network.
    First, observe that the assumption of \Cref{lem:homogenization_function} is satisfied for all~$x\in\R^k$ by our constructed network.
    Further, we have $h(x,0)=0$ for all $x\in \R^k$ since the deletion of biases in all $p_{c,i}$ terms leads to the cancellation of all terms in every~$\delta_D$ function in the second hidden layer.
    Thus, by the same argument as in the proof of \Cref{thm:positivity_hardness_without_bias}, we conclude that $h$ has a positive point if and only if $f$ has a positive point.

    \textbf{Case II: Odd $\ell \ge 3$.} We reduce from the~W[$\ell$]-hard \textsc{Weighted Antimonotone $\ell$-Normalized SAT} problem \citep{DF13}.
    The reduction is very similar to Case~I, except that we need to adjust the first two hidden layers in order to enforce that $k$ variables are set to true by the input and to compute a conjunction of negative literals.
    
    Given an instance~$(F,k)$ with an antimonotone $\ell$-normalized formula~$F$ with variables $z_1,\ldots,z_n$, we construct a ReLU network with~$\ell+1$ layers and input dimension~$k$.
    
    In the first hidden layer, we use the same neurons computing functions~$p_{c,i}$ as in the construction for Case~I above.
    Additionally, for each~$c\in[k]$, we add $3n$ neurons computing the function $p_c\colon\R^k\to[0,1]$ with 
    \begin{equation*}
    p_c(x)\coloneqq\sum_{i\in [n]}\big(\max(0, 8(x_c-i+\frac{1}{8}))
    -\max(0, 16(x_c-i))
    +\max(0, 8(x_c-i-\frac{1}{8}))\big),
    \end{equation*}
    which satisfies $p_c(x)=1 \iff x_c \in [n]$ (analogous to \Cref{fig:penalty_function}, replace $\omega_{c,i}$ with $i$).
    Moreover, for each pair~$c\neq c'\in[k]$, we add 3 neurons computing the function $s_{c,c'}\colon\R^k\to[0,1]$ with
    \[s_{c,c'}(x)\coloneqq 1-\max(0,x_c-x_{c'}+1)+\max(0,2(x_c-x_{c'}))-\max(0,x_c-x_{c'}-1),\]
    which satisfies $s_{c,c'}(x)=\min(1,|x_c-x_{c'}|)$ (see \Cref{fig:scc_plot}).
    The idea is that these functions enforce that the~$k$ input values correspond to~$k$ pairwise different variable indices.

\begin{figure}
    \centering
    \includegraphics[width=0.275\textwidth, page=14]{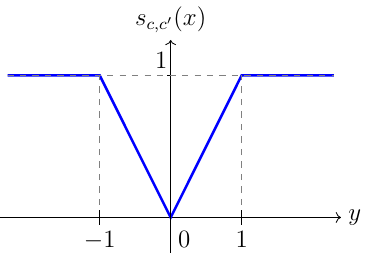}
    \hspace{0.75cm}
    \includegraphics[width=0.22\textwidth, page=15]{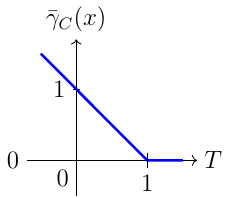}
    \hspace{0.75cm}
    \includegraphics[width=0.275\textwidth, page=16]{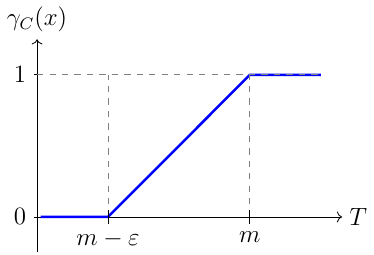}
    \caption{Plot of $s_{c,c'}$ with $y=x_c-x_{c'}$ (left), $\bar{\gamma}_C$ with $T=\sum_{c\in[k]}\sum_{j=1}^m p_{c,i_j}(x)$ for an innermost conjunction (center) and~$\gamma_C$ for other conjunctions with $T=\sum_{j=1}^m \delta_{D_j}$ (right).}
    \label{fig:scc_plot}
\end{figure}
    
    The second hidden layer then simulates an antimonotone conjunction $C=\bar{z}_{i_1}\wedge \cdots\wedge \bar{z}_{i_m}$ via
    \[\bar{\gamma}_C(x)\coloneqq \max(0,1-\sum_{c\in[k]}\sum_{j=1}^mp_{c,i_j}(x))\in[0,1].\]
    Note that $\bar{\gamma}_C(x)=0$ if some~$p_{c,i_j}(x)=1$ (see \Cref{fig:scc_plot}).

    Now, the remaining $\ell-1$ layers simulate the disjunctions and conjunctions in~$F$ almost in the same way as in Case~I above, except that we modify $\gamma_C$ for conjunctions:
    Given an~$\varepsilon\in (0,1)$ (which we will make explicit later), we simulate a conjunction $C=D_1\wedge \cdots \wedge D_m$ with
    \[
    \gamma_C(x)\coloneqq
    \max\left(0,\frac{1}{\varepsilon}\left(\sum_{j=1}^m \delta_{D_j}(x)-m+\varepsilon\right)\right)
    -
    \max\left(0,\frac{1}{\varepsilon}\left(\sum_{j=1}^m \delta_{D_j}(x)-m\right)\right)\in [0,1].
    \]
    Note that $\gamma_C(x)>0$ if and only if $\sum_{j=1}^m \delta_{D_j}(x)> m-\varepsilon$ (see \Cref{fig:scc_plot}).
    In particular, if $\gamma_C(x)>0$ then $\delta_{D_j}(x)>1-\varepsilon$ for all $j\in[m]$.
    
    Moreover, the values of the $p_c$ and~$s_{c,c'}$ functions from the second layer are propagated to the output neuron (via the identity function which can be realized with two neurons).
    The output neuron then computes
    \[f(x)\coloneqq\sum_{j=1}^m\delta_{D_j}(x)+\sum_{c=1}^kp_c(x)+\sum_{c< c'}s_{c,c'}(x)-m-k-\binom{k}{2}+\varepsilon,\]
    with $\varepsilon\coloneqq \frac{1}{M+1}>0$, where $M\ge 1$ is the maximum fan-in for any disjunction in $F$.
    Note that $\varepsilon \leq \frac{1}{2}$ and thus $\varepsilon/ 4\leq 1-\varepsilon$.
    
    The correctness is analogous to Case~I.
    Let $i_1,\ldots,i_k$ be the indices of the variables which are set to true to satisfy~$F$. Then, to see that $f(x^*)>0$ holds for $x^*\coloneqq(i_1,\ldots,i_k)$ note that $p_j(x^*)=p_{j,i_j}(x^*)=1$ for every~$j\in[k]$ and also~$s_{j,j'}(x^*)=1$ for all~$j\neq j'\in[k]$. Therefore, in the second layer, $\bar{\gamma}_C(x^*)=1$ holds for each innermost conjunction~$C$ of~$F$ which is satisfied.
    Consequently, in the next layer $\delta_D(x^*)=1$ holds for every following disjunction~$D$ in~$F$ which is satisfied.
    Hence, it inductively follows that $f(x^*)=\varepsilon>0$ since~$F$ is satisfied and $\sum_{c=1}^kp_c(x^*)+\sum_{c< c'}s_{c,c'}(x^*)=k+\binom{k}{2}$.

    Conversely, if $f(x^*)>0$ for $x^*\in \R^k$, then let $I\coloneqq \{i\in [n]\mid \exists c\in [k]: p_{c,i}(x^*)>1-\varepsilon\}$.
    We claim that $|I|=k$.
    To see this, note that $f(x^*)>0$ implies $p_c(x^*) >  1-\varepsilon$ for all $c$, $s_{c,c'}(x^*)>1-\varepsilon$ for all $c< c'$, and 
    $\delta_{D_j}(x^*)> 1-\varepsilon$ for all $j$.
    Next, observe that $p_c(x^*) >  1-\varepsilon$ implies that there is a unique index $i_c\in [n]$ such that $x^*_c\in (i_c-\varepsilon/8,i_c+\varepsilon/8)$ with $p_{c,i_c}(x^*)=p_c(x^*)$ and $p_{c,i'}(x^*)=0$ for all other $i'\neq i_c$.
    Hence $I=\{i_1,\dots,i_k\}$.
    To see that there are no duplicates, observe that $s_{c,c'}(x^*)>1-\varepsilon$ is equivalent to $|x^*_c-x^*_{c'}|>1-\varepsilon$.
    If $i_c=i_{c'}$ holds, then $x^*_c$ and $x^*_{c'}$ are both contained in $(i_c-\varepsilon/8,i_c+\varepsilon/8)$ and $|x^*_c-x^*_{c'}|<\varepsilon/4\leq 1-\varepsilon$ contradicts the assumption that $s_{c,c'}(x^*)>1-\varepsilon$.
    Hence, $i_c\neq i_{c'}$ holds and thus $|I|=k$.

    Now, we claim that setting the variables $z_i$ with $i\in I$ to true gives a weight-$k$ satisfying assignment for $F$.
    To verify this, note that $\delta_{D_j}(x^*)>1-\varepsilon$ with $D_j=C_{j,1}\lor \cdots \lor C_{j,m_j}$ implies $\sum_{i=1}^{m_{j}}\gamma_{C_{j,i}}(x^*)>1-\varepsilon$,
    which holds only if $\gamma_{C_{j,i}}(x^*)>(1-\varepsilon)/m_j$ holds for at least one $i$.
    Recall that $\gamma_{C_{j,i}}(x^*)>0$ implies that all disjunction functions $\delta_D$ of $C_{j,i}$ satisfy $\delta_D(x^*)>1-\varepsilon$.
    Inductively, we arrive at a disjunction $D=C_1\lor \cdots \lor C_m$ with $C_j=\bar{z}_{j_1}\wedge \cdots\wedge \bar{z}_{j_t}$ such that there is at least one $j\in [m]$ with $\bar{\gamma}_{C_j}(x^*)>(1-\varepsilon)/m$ (so $D$ is one of the innermost disjunctions).
    Note that if $C_j$ is satisfied by the assignment corresponding to $I$, then by the same inductive argument as in Case~I, this assignment satisfies the overall formula $F$.
    Assume for the sake of contradiction that $C_j$ is not satisfied by the assignment corresponding to $I$.
    Thus, for some $l^*\in[t]$, the literal $\bar{z}_{j_{l^*}}$ must evaluate to false, which means that $j_{l^*}=i_c\in I$ for some $c\in[k]$, since all variables set to true in this assignment are of the form $z_{i_c}$ for some $c\in [k]$.
    Therefore, we have $p_{c,j_{l^*}}(x^*)=p_{c,i_c}(x^*)>1-\varepsilon$ which gives
    \[
    \bar{\gamma}_{C_j}(x^*)=\max(0, 1-\sum_{l=1}^t \sum_{c\in [k]}p_{c,j_l}(x^*))\leq 1-p_{c,j_{l^*}}(x^*)<\varepsilon,
    \]
    contradicting the fact that $\bar{\gamma}_{C_j}(x^*)>(1-\varepsilon)/m = M/((M+1)m)=\varepsilon M/m \ge \varepsilon$.
    Thus, $C_j$ is satisfied by $I$, which proves that $I$ yields a weight-$k$ satisfying assignment for $F$.
    
    \textbf{Removing biases in Case II.}
    Let $h\colon\R^{k+1}\to \R$ be the homogenization (see \Cref{def:homogenization}) of the network constructed in Case~II.
    First, observe that the assumption of \Cref{lem:homogenization_function} is satisfied for all $x\in\R^k$ by the construction.
    Further, we have $h(x,0)=0$ for all $x\in \R^k$ since the deletion of biases in all $p_{c,i}$, $p_c,$ and $s_{c,c'}$ terms leads to the cancellation of all terms in every~$\bar{\gamma}_C$ function and the function of each of the two extra neurons in the second hidden layer.
    Thus, by the same argument as in the proof of \Cref{thm:positivity_hardness_without_bias}, we conclude that $h$ has a positive point if and only if $f$ has a positive point.
\end{proof}

\Cref{thm:Wl-hardness} also implies W[$\ell$]-hardness w.r.t.\ the input dimension $d$ for approximating the maximum of a $(\ell +1)$-layer ReLU network over a polyhedron within any multiplicative factor for every fixed $\ell\geq 1$.
\citet[Corollary 13]{FGS25} showed that approximating this value is NP-hard.

\begin{corollary}\label{thm:max_approx_hardness}
    For every fixed $\ell\geq 1$, approximating the maximum of an $(\ell+1)$-layer ReLU network over a polyhedron given in halfspace representation within any multiplicative factor is \textnormal{W[$\ell$]}-hard with respect to its input dimension $d$ and cannot be done in $\rho(d)\cdot N^{o(d)}$ time (where $N$ is the combined encoding size of the network and the polyhedron) for any function $\rho$ assuming the ETH.
\end{corollary}

\begin{proof}
    In \Cref{thm:Wl-hardness}, for every $\ell\geq 1$, we constructed an $(\ell+1)$-layer network computing $f\colon\R^d\to\R$ for which it is W[$\ell$]-hard to distinguish whether $\max_{x\in [0,M]^d}f(x) > 0$ or $\max_{x\in [0,M]^d}f(x) \leq 0$ holds, where $M>0$ is a suitable value whose encoding size is polynomial in the overall encoding length of the network: for $\ell=1$, where we reduce from \textsc{Multicolored Clique}, take $M=|V|^3$, and for $\ell \geq 2$, where we reduce from \textsc{Weighted (Anti)Monotone $t$-Normalized SAT}, take $M=n+1$.
    Thus, approximating $\max_{x\in [0,M]^d}f(x)$ within any multiplicative factor would allow us to distinguish between the two cases.
\end{proof}

By adding another hidden layer with a single ReLU neuron to the networks constructed in the proof of \Cref{thm:Wl-hardness}, we obtain for every fixed $\ell\geq 1$ an $(\ell+2)$-layer ReLU network that has a non-zero output if and only if the original $(\ell+1)$-layer network has a positive output. This yields the following corollary.

\begin{corollary}\label{cor:hardness_zero_check_3layers}
    For every fixed $\ell\geq 1$, the problem of deciding whether an $(\ell+2)$-layer ReLU network computes a non-zero function is \textnormal{W[$\ell]$}-hard with respect to its input dimension $d$ and not solvable in $\rho(d)\cdot N^{o(d)}$ time (where $N$ is the input bit-length) for any function $\rho$ assuming the ETH.
\end{corollary}

The NP-hardness of deciding the non-zero function for 3 layers was established by \citet{FGS25}.
For 2 layers, it is solvable in polynomial time \citep{FGS25}, which holds also in the presence of biases \citep{stargalla2025computational}.
Thus, \Cref{cor:hardness_zero_check_3layers} draws an even sharper boundary between the computational complexity of this problem in the 2-layer and 3-layer cases.

Observe that \Cref{cor:hardness_zero_check_3layers} implies an even more general hardness result:
For every family of functions $(f_d)_{d\in \N}$ where $f_d\colon\R^d\to \R$ can be computed by a ReLU network with $\ell\ge 3$ layers that can be constructed in $\poly(d)$ time, it is coNP- and coW[$\ell-2$]-hard with respect to~$d$ to decide whether a given $\ell$-layer ReLU network with input dimension $d$ computes $f_d$.
The reduction simply adds the $\ell$-layer network from \Cref{cor:hardness_zero_check_3layers} to the network computing $f_d$.

We obtain two further corollaries.
First, observe that in the case where the function from \Cref{cor:hardness_zero_check_3layers} is non-zero, it is not convex, as it is bounded from below by $0$ and from above by~$1$.
This implies hardness of deciding whether an $(\ell+2)$-layer ReLU network computes a convex function for any $\ell\geq 1$.
In contrast, this problem is polynomial-time solvable in the 2-layer case. In this case, the points of non-differentiability form a hyperplane arrangement and the function is convex if and only if it is convex around each of these hyperplanes. Hence, convexity can be checked locally at the hyperplanes: for each hyperplane, one groups all neurons whose preactivation vanishes on it and checks whether their total contribution is convex. Since these groups can be found by comparing proportional weight vectors, this gives a polynomial-time algorithm.
\begin{corollary}\label{cor:hardness_convexity_check}
    For every fixed $\ell\geq 1$, deciding whether an $(\ell+2)$-layer ReLU network computes a convex function is \textnormal{W[$\ell]$}-hard with respect to its input dimension $d$ and not solvable in $\rho(d)\cdot N^{o(d)}$ time (where $N$ is the input bit-length) for any function $\rho$ assuming the ETH.
\end{corollary}

Further, observe that in the case where the function from \Cref{cor:hardness_zero_check_3layers} is non-zero for $\ell=1$, the function computed by this 3-layer ReLU network is not computable by a 2-layer ReLU network, as it is zero everywhere outside a sufficiently large box around the origin and thus its points of non-differentiability do not form a hyperplane arrangement.
This leads to the following hardness result.
\begin{corollary}\label{cor:hardness_2-layer_check}
    Deciding whether a $3$-layer ReLU network computes a function representable by a $2$-layer ReLU network is W[1]-hard with respect to its input dimension $d$ and not solvable in $\rho(d)\cdot N^{o(d)}$ time (where $N$ is the input bit-length) for any function $\rho$ assuming the ETH.
\end{corollary}

\citet{FGS25} proved NP-hardness of \textsc{2-Layer ReLU Surjectivity} and asked whether the problem is fixed-parameter tractable with respect to $d$.
We give a negative answer to this question.

\begin{theorem}\label{thm:surjectivity_Whardness}
    For every fixed $\ell\geq 1$, \textsc{$(\ell+1)$-Layer ReLU Surjectivity} is \textnormal{W[$\ell$]}-hard with respect to $d$ and not solvable in $\rho(d)\cdot N^{o(d)}$ time (where $N$ is the input bit-length) for any function $\rho$ assuming the ETH.
\end{theorem}
\begin{proof}
    Recall that a positively homogeneous function $g \colon \R^d \to \R$ is surjective if and only if there exist two points $v^+,v^- \in \R^d$ such that $g(v^+) > 0$ and $g(v^-)<0$.
    Note that for every fixed $\ell\geq 1$, the function $f\colon\R^{k+1}\to \R$ computed by the $(\ell+1)$-layer ReLU network from the proof of \Cref{thm:Wl-hardness} is positively homogeneous and satisfies $f(0,1)<0$.
    Thus, $f$ is in fact surjective if and only if it has a positive point.
\end{proof}

\begin{figure}
\centering
\includegraphics[width=0.8\textwidth, page=17]{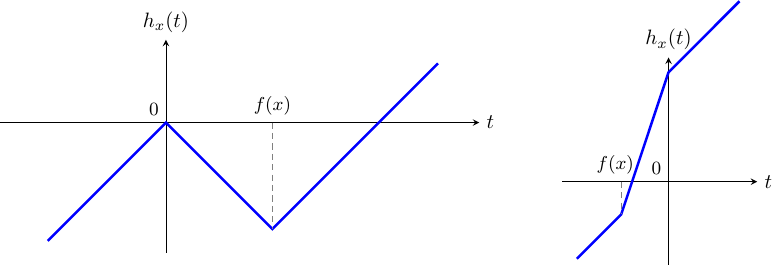}
\caption{Sketch of the function $h_x(t)=t-2\max(0,t)+2\max(0,t-f(x))$ for $f(x)>0$ (left) and $f(x)\leq 0$ (right).
If $f(x)>0$, then $h_x$ is not injective.
If $f(x)\leq 0$, then $h_x$ is strictly increasing and thus injective.}
\label{fig:injectivity_hx_function}
\end{figure}

\citet{FGS25} showed that deciding injectivity of 2-layer ReLU networks with~$n$ hidden neurons can be done in $(d+1)^d\poly(N)$ time and thus the problem is FPT for the input dimension~$d$.
We show that the complexity of injectivity increases with the number of hidden layers.
\begin{theorem}\label{thm:injectivity_Whardness}
    For every fixed $\ell\geq 1$, deciding whether an $(\ell+2)$-layer ReLU network is not injective is \textnormal{W[$\ell$]}-hard with respect to $d$ and not solvable in $\rho(d)\cdot N^{o(d)}$ time (where $N$ is the input bit-length) for any function $\rho$ assuming the ETH.
\end{theorem}
\begin{proof}
    Fix $\ell\geq 1$, and let $f\colon\R^d\to \R$ be the function computed by the $(\ell+1)$-layer network constructed in the proof of \Cref{thm:Wl-hardness}.
    We define $g\colon\R^{d+1}\to \R^{d+1},\,(x,t)\mapsto(x,h(x,t))$, where
    \[
    h\colon\R^{d+1}\to \R, \; (x,t)\mapsto t-2\max(0,t)+2\max(0,t-f(x)).
    \]
    The function $g$ is computed by an $(\ell+2)$-layer network obtained as follows.
    First, we can easily build a 2-layer ReLU network computing the function $h$ where $t$ and $f(x)$ are treated as one-dimensional inputs, and then replacing the ``input'' $f(x)$ by the $(\ell+1)$-layer network computing $f$.
    For each fixed $x\in \R^d$, let $h_x\colon\R\to \R,\, t\mapsto h(x,t)$.
    We first observe that $g$ is injective if and only if $h_x$ is injective for every $x\in \R^d$.
    To see this, suppose that $g(x,t)=g(x',t')$.
    By the definition of $g$, we have $x=x'$ and thus $h_x(t)=h(x,t)=h(x,t')=h_x(t')$.
    Hence, if $h_x$ is injective for every $x\in \R^d$, then $t=t'$, and therefore $g$ is injective.
    Conversely, if $h_x$ is not injective, then there exist $t\neq t'$ with $h_x(t)=h_x(t')$.
    It follows that $g(x,t)=(x,h_x(t))=(x,h_x(t'))=g(x,t')$ and hence $g$ is not injective.
    It remains to characterize when $h_x$ is injective.
    If $f(x)\leq 0$, then $h_x$ is strictly increasing and thus injective.
    If $f(x)>0$, then $h_x$ is not injective, as $h_x(-f(x))=-f(x)=h_x(f(x))$.
    Thus, $h_x$ is injective if and only if $f(x)\leq 0$ (see \Cref{fig:injectivity_hx_function}).
    Therefore, $g$ is not injective if and only if there is an $x\in \R^d$ with $f(x)> 0$.
    Thus, deciding non-injectivity of $g$ is at least as hard as deciding positivity of the network computing $f$.
    The W[$\ell$]-hardness and ETH-based running time lower bound therefore follow from \Cref{thm:Wl-hardness}.
\end{proof}

\section{Hardness of Zonotope Non-Containment}
\label{sec:zonotope_containment_hardness}
In this section, we prove W[1]-hardness for \textsc{Zonotope Non-Containment}, the complement of \textsc{Zonotope Containment}.
\textsc{Zonotope Containment} is coNP-complete, and can be solved in $n^{d-1}\cdot \poly(N)$ time (where $N$ is the input bit-length) by enumerating the vertices of one zonotope, but fixed-parameter tractability with respect to the dimension $d$ remained open so far \citep{FGHS25}.  Moreover, \citet{KULMBURG202184}  showed that containment is equivalent to maximizing a certain zonotope norm, making it a special case of norm maximization on zonotopes.

\begin{figure}
\centering
\includegraphics[width=0.9\textwidth, page=18]{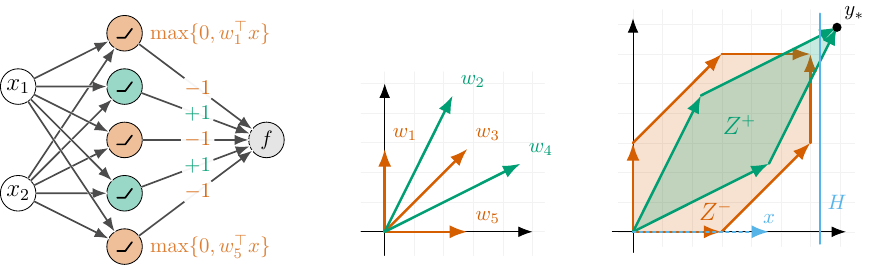}
\caption{An illustration of the equivalence between \textsc{2-layer ReLU Positivity} and \textsc{Zonotope Non-Containment}.
The weights of a bias-free ReLU network (left and center) induce the zonotopes $Z^+$ and $Z^-$ (right).
Let $H=\{y \in \R^d \mid y^\top x = b\}$ be a hyperplane that separates $y_*\in Z^+$ from $Z^-$. Then $g(x) =\max_{y \in Z^+} y^\top x - \max_{y \in Z^-} y^\top x > y_*^\top x - b>0$.}
\label{fig:zonotope_containment}
\end{figure}

\citet{FGS25} showed that \textsc{2-layer ReLU Positivity} (without biases) is equivalent to \textsc{Zonotope Non-Containment} following from the duality of positively homogeneous convex CPWL functions from $\R^d$ to $\R$ and polytopes in $\R^d$. We briefly sketch this equivalence.
Let a $2$-layer ReLU network without biases be given by $g(x) = \sum_{i=1}^m \lambda_i \max\{0,w_i^\top x\}$. We can assume without loss of generality that $\lambda_i \in \{-1,1\}$ due to the positive homogeneity of $\max\{0,w_i^\top x\}$. Hence, $g(x)=\sum_{i\in I^+}\max\{0,w_i^\top x\} - \sum_{i\in I^-}\max\{0,w_i^\top x\}$ for vectors $w_i\in \R^{d},i\in I^+\cup I^-$. Defining the zonotopes
\[
Z^+\coloneqq \varphi(\sum_{i\in I^+}\max\{0,w_i^\top x\}) = \sum_{i\in I^+}\conv(\{0,w_i\}),
\]
\[
Z^- \coloneqq \varphi(\sum_{i\in I^-}\max\{0,w_i^\top x\})=\sum_{i\in I^-}\conv(\{0,w_i\}),
\]
it holds that  $g = \varphi^{-1}(Z^+) -  \varphi^{-1}(Z^-)$. By definition of the support function, $Z^+\subseteq Z^-$ implies $\varphi^{-1}(Z^+) \leq \varphi^{-1}(Z^-)$. Conversely, if $y_* \in Z^+ \setminus Z^-$, then there is a separating hyperplane $H=\{y \in \R^d : y^\top x =b\}$  such that $y^\top_* x > b$ and $y^\top x < b$ for all $y \in Z^-$ (see \Cref{fig:zonotope_containment}). Hence, $g(x) > 0$.
Since also any pair of zonotopes is of the form above, \textsc{2-layer ReLU Positivity} without biases is equivalent to \textsc{Zonotope Non-Containment}.
Thus, \Cref{thm:positivity_hardness_without_bias} implies the following theorem.

\begin{theorem}
\label{thm:zonotope-cont-hardness}
    \textsc{Zonotope Non-Containment} is \textnormal{W[1]}-hard with respect to $d$ and not solvable in $\rho(d)\cdot N^{o(d)}$ time (where $N$ is the input bit-length) for any function $\rho$ assuming the ETH.
\end{theorem}

\begin{proof}
    Follows from \Cref{thm:positivity_hardness_without_bias} and the equivalence to \textsc{2-Layer ReLU Positivity} without biases \citep[Proposition 18]{FGS25}.
\end{proof}

\section{Complexity of Computing the Lipschitz Constant}\label{sec:lipschitz_hardness}
\citet{JD20} established the NP-hardness for approximating the $L_p$-Lipschitz constant for $p=1$ and $p=\infty$ for 3-layer ReLU networks within a multiplicative factor of $\Omega(N^{1-\varepsilon})$ for every constant $\varepsilon>0$, where $N$ is the encoding size of the ReLU network.
The NP-hardness result of \citet{FGS25} for \textsc{2-layer ReLU Positivity} implies NP-hardness for approximating the $L_p$-Lipschitz constant for $p\in(0,\infty]$ within any multiplicative factor for 3-layer ReLU networks.
We extend this by showing W[1]-hardness of the problem.

\begin{corollary}\label{cor:hardness_lipschitz_approx_3layers}
    For every fixed $\ell\geq 1$ and $p\in (0,\infty]$, approximating the $L_p$-Lipschitz constant of an $(\ell+2)$-layer ReLU network by any multiplicative factor is \textnormal{W[$\ell$]}-hard with respect to its input dimension $d$ and cannot be done in $\rho(d)\cdot N^{o(d)}$ time (where $N$ is the input bit-length) for any function $\rho$ assuming the ETH.
\end{corollary}
\begin{proof}
    For every fixed $\ell\geq 1$, adding a hidden layer with a single ReLU neuron to the construction in the proof of~\Cref{thm:Wl-hardness} yields an $(\ell+2)$-layer network which computes a function with a non-zero~$L_p$-Lipschitz constant if and only if the original $(\ell+1)$-layer network has a positive output.
    Hence, any multiplicative approximation could be used to decide \textsc{$(\ell+1)$-layer ReLU $L_p$-Lipschitz Constant}.
    \end{proof}
    
Next, we extend our result to exact computation.
For this, we use the following well-known statement on the Lipschitz constant of CPWL functions.
For the sake of completeness, we add an elementary proof.
\begin{lemma}\label{lem:lipschitz_region}
    Let $f\colon\R^d\to \R$ be a CPWL function and let $\mathcal{C}$ be a set of full-dimensional pairwise disjoint polyhedra such that $\bigcup_{C\in \mathcal{C}}C=\R^d$ and $f(x)=a_C^\top x +b_C$ for all $x\in C$ and some~$a_C\in\R^d$, $b_C\in\R$.
    Then, for all $p\in (0,\infty]$, we have
    \[
    L_p(f)=\max_{C\in \mathcal{C}}\max_{\|x\|_p\leq 1}|a_C^\top x|=\max_{C\in \mathcal{C}}\|a_C\|_q,
    \]
    where $\|\cdot\|_q$ is the dual norm of the $L_p$-norm for $p\in[1,\infty]$ and the infinity norm for $p\in (0,1)$.
\end{lemma}
\begin{proof}
    Fix $p\in (0,\infty]$ and define $\mathcal{L}_p\coloneq \max_{C\in \mathcal{C}}\max_{\|x\|_p\leq 1}|a_C^\top x|$.
    We will show $L_p(f)=\mathcal{L}_p$.

    First, fix $x\in \text{int}(C)$ for some $C\in \mathcal{C}$.
    For all $h\in \R^d$, there is a $t>0$ with $x+th\in C$.
    Thus
    \[
    L_p(f)\geq\dfrac{|f(x+th)-f(x)|}{\|th\|_p}
    =\dfrac{|a_C^\top (th)|}{\|th\|_p}
    =\left|a_C^\top (\frac{h}{\|h\|_p})\right|
    \quad\text{for all }h\neq0,
    \]
    which yields $L_p(f)\geq \mathcal{L}_p$ since $C$ was arbitrary.
    
    For the other direction, fix $x,y\in \R^d,\,x\neq y$ and set $h=y-x$.
    Let $0=t_0<t_1<\dots <t_k=1$ be the points where $t\mapsto f(x+th)$ is non-differentiable, that is, $\{x+th: t\in[t_{j-1},t_j]\}\subseteq C_j$ holds for every $j\in [k]$.
    Applying the triangle inequality and $|a_{C_j}^\top h|\leq \mathcal{L}_p \|h\|_p$ yields
    \[
    |f(y)-f(x)|
    \leq \sum_{j=1}^k|f(x+t_jh)-f(x+t_{j-1}h)|
    = \sum_{j=1}^k(t_j-t_{j-1})|a_{C_j}^\top h|
    \leq \mathcal{L}_p \|h\|_p,
    \]
    so $\frac{|f(y)-f(x)|}{\|y-x\|_p}\leq \mathcal{L}_p$.
    Since this holds for all $x\neq y$, we obtain $L_p(f)\leq\mathcal{L}_p$.

    What remains is to show that $\mathcal{L}_p=\max_{C\in \mathcal{C}}\|a_C\|_q$.
    For $p\in[1,\infty]$, it is well known that $\|\cdot\|_p$ defines a norm and that $\max_{\|x\|_p\leq 1}|a_C^\top x|=\|a_C\|_q$ holds for any $a_C\in \R^d$, where $\|\cdot\|_q$ is the dual norm of the $L_p$ norm, which implies the equality.

    Now, consider the case where $p\in(0,1)$.
    To see that $\max_{\|x\|_p\leq 1}|a_C^\top x|=\|a_C\|_\infty$, observe that for any $x\in \R^d$ with $\|x\|_p\leq 1$ (and thus also for $|x_i|\leq 1$ for all $i\in [d]$), we have
    \[
    |a_C^\top x|
    \leq \sum_{i=1}^{d}|(a_C)_i|\cdot|x_i|
    \leq \|a_C\|_\infty\sum_{i=1}^{d}|x_i|^p
    \leq \|a_C\|_\infty\|x\|_p^p
    \leq \|a_C\|_\infty.
    \]
    For the other direction, choose an index $j\in [d]$ with $|(a_C)_j|=\|a_C\|_\infty$.
    Then, we have $\|a_C\|_\infty=|a_C^\top e_j|\leq \max_{\|x\|_p\leq 1}|a_C^\top x|$, which proves the equality.
\end{proof}

\citet{VS18} showed NP-hardness for \textsc{2-Layer ReLU $L_2$-Lipschitz Constant}.
We extend the NP-hardness to $p\in(0,\infty]$ and show W[1]-hardness w.r.t. $d$.

\begin{theorem}\label{thm:lipschitz_constant_exact_hardness}
    For all $p\in (0,\infty]$, \textsc{$2$-Layer ReLU $L_p$-Lipschitz Constant} is \textnormal{NP}-hard, \textnormal{W[1]}-hard with respect to $d$ and not solvable in $\rho(d)\cdot N^{o(d)}$ time (where $N$ is the input bit-length) for any function $\rho$ assuming the ETH.
\end{theorem}
\begin{proof}
    We give a parameterized (polynomial-time) reduction from \textsc{Multicolored Clique} to \textsc{$2$-Layer ReLU $L_p$-Lipschitz Constant}.
    
    Let $(G=(V=V_1\dot\cup\cdots\dot\cup V_k, E),k)$ be an instance of \textsc{Multicolored Clique}.
    Further, let $f\colon\R^{k}\to \R$ be the function computed by the network constructed in the proof of \Cref{lem:main_reduction}.
    Set $T\coloneq k+\binom{k}{2}$.
    Recall that $\max_{x\in \R^k}f(x)=T$ if and only if $G$ has a $k$-colored clique and otherwise on each input, at least one $p_c$ or $s_{r,l}$ function outputs zero.
    The idea is to use this fact to create a separation of the Lipschitz constant between these cases.

    We now introduce an additional variable $y$ and define a 2-layer ReLU network computing
    \[
    h\colon\R^{k+1}\to \R,\quad h(x,y)
    \coloneqq f(x+Ty (1,\dots, 1)^\top)
    =\sum_{c\in [k]}p_c(x_c+Ty)+\sum_{(r,l)\in \binom{[k]}{2}}s_{r,l}(x_r+Ty,x_l+Ty).
    \]
    On every full-dimensional linear region $R\in \Sigma_h$ of $h$, we have $h(x,y)=a_R^\top (x,y)+b_R$ for all $(x,y)\in R$.
    The definition of $p_c$ and $s_{r,l}$ imply
    \[
    a_R=\sum_{c\in [k]}\alpha_{c,R}(e_c+Te_y)+\sum_{(r,l)\in \binom{[k]}{2}}\beta_{r,l,R}(e_r+e_l+2Te_y),
    \]
    where $\alpha_{c,R}\in \{-8,0,8\}$ and $\beta_{r,l,R}\in \{-4,0,4\}$.
    Note that the coefficients $\alpha_{c,R}$ and $\beta_{r,l,R}$ are nonzero if and only if the corresponding node and edge gadget is strictly positive on $R$, respectively.
    
    First, suppose that $G$ has a $k$-colored clique $\{v_{1,a_1},\dots, v_{k,a_k}\}$.
    Consider the point $(x^*,0)$ with $x^*_c\coloneqq\omega_{c,a_c}-1/16$ for all $c\in [k]$.
    One can easily check that this point does not lie on any hyperplane induced by the neurons of $h$.
    Thus, it lies in the interior of a full dimensional linear region $R^*$ with
    \[
    a_{R^*}=\sum_{c\in [k]}8(e_c+Te_y)+\sum_{(r,l)\in \binom{[k]}{2}}4(e_r+e_l+2Te_y)
    \]
    and thus $(a_{R^*})_y=8T^2$.
    Since $\|e_y\|_p=1$ for every $p\in (0,\infty]$, we have by \Cref{lem:lipschitz_region}
    \[
    L_p(h)=\max_{R\in \Sigma_h}\max_{\|(x,y)\|_p\leq 1}|a_R^\top (x,y)|\geq |a_{R^*}^\top e_y|=8T^2.
    \]
    Now, suppose that $G$ does not have a $k$-colored clique.
    We first consider the case $p\in [1,\infty]$.
    Fix $R\in \Sigma_h$.
    Since at least one gadget has output zero, the triangle inequality yields
    \begin{align*}
    \|a_R\|_q
    &\leq \sum_{c\in [k]}|\alpha_{c,R}|(\|e_c\|_q+T\|e_y\|_q)+\sum_{(r,l)\in \binom{[k]}{2}}|\beta_{r,l,R}|(\|e_r\|_q+\|e_l\|_q+2T\|e_y\|_q)\\
    &
    \leq \sum_{c\in [k]}|\alpha_{c,R}|(T+1)+\sum_{(r,l)\in \binom{[k]}{2}}2|\beta_{r,l,R}|(T+1)\\
    &\leq 8(T+1) \cdot \left(\left|\{c\in [k]: \alpha_{c,R}\neq 0\}\right|+\left|\{(r,l)\in \binom{[k]}{2}: \beta_{r,l,R}\neq 0\}\right|\right)\\
    &\leq 8(T+1)(T-1)
    \end{align*}
    where $\|\cdot\|_q$ is the dual norm of the $L_p$-norm.
    Now, consider the case $p\in (0,1)$.
    By \Cref{lem:lipschitz_region}, we have $\max_{\|(x,y)\|_p\leq 1}|a_R^\top (x,y)|=\|a_R\|_\infty$ and in the same way as in the case of $p\in [1,\infty]$, we obtain $\|a_R\|_\infty\leq 8T^2-8$.
    Hence
    \[
    L_p(h)
    =\max_{R\in \Sigma_h}\max_{\|(x,y)\|_p\leq 1}|a_R^\top (x,y)|
    =\max_{R\in \Sigma_h}\|a_R\|_q
    \leq 8(1+T)(T-1)
    =8T^2-8,
    \]
    which implies a separation for the $L_p$-Lipschitz constant $L_p(h)$ depending on whether $G$ has a $k$-colored clique or not.
    With this, the \textsc{$2$-Layer ReLU $L_p$-Lipschitz Constant} instance consisting of $L=8T^2$ and the underlying network of $h$ is a yes-instance if and only if $(G,k)$ is a yes-instance of \textsc{Multicolored Clique}.
\end{proof}

Using similar ideas, we can extend the hardness also to W[$\ell$]-hardness for $\ell+1$ layers.

\begin{theorem}\label{thm:Wl-hardness-Lipschitz}
For every fixed~$\ell\ge 1$ and for all $p\in (0,\infty]$, \textsc{$(\ell+1)$-Layer ReLU $L_p$-Lipschitz Constant} is \textnormal{W[$\ell$]}-hard with respect to~$d$ and not solvable in $\rho(d)\cdot N^{o(d)}$ time (where $N$ is the input bit-length) for any function $\rho$ assuming the ETH.
\end{theorem}
\begin{proof}
The case $\ell=1$ is \Cref{thm:lipschitz_constant_exact_hardness}.

Let $\ell\geq 2$ be fixed.
We give a polynomial-time parameterized reduction from the~W[$\ell$]-hard \textsc{Weighted (Anti)Monotone $\ell$-Normalized SAT} problem.
Given an instance $(F,k)$, observe that the $(\ell+1)$-layer network constructed in the proof of \Cref{thm:Wl-hardness} computes $\sum_{i=1}^m U_i(x) + B$, where $U_i$ is either $\delta_{D}$, a $p_c$, or a $s_{c,c'}$ term, depending on whether $\ell$ is even or odd, and $B$ is a bias term.
Recall from the proof of \Cref{thm:Wl-hardness} that, if the instance is a yes-instance, there is a point $x^*$ with $U_i(x^*)>T$ for all $i\in [m]$, where $T$ is a threshold value with $T=0$ in the even and $T=1-\varepsilon$ in the odd case.
Conversely, if the instance is a no-instance, for every $x\in \R^k$, there is some $i\in[m]$ with $U_i(x)\leq T$.
We use this to create a separation of the Lipschitz constant between these cases.

We add an input variable $y$ and construct an $(\ell+1)$-layer network computing
\[
h:\R^{k+1}\to \R,\quad h(x,y)=\sum_{i=1}^m (\max(0,V_i(x)-T)-\max(0,V_i(x)-T-U\max(0,y))),
\]
where $V_i(x)=p_c(x)$ if $U_i=p_c$, $V_i(x)=s_{c,c'}(x)$ if $U_i=s_{c,c'}$, $V_i(x)=\sum_{j=1}^{m'} \gamma_{C_j}(x)$ if $U_i=\delta_D$ and $\ell\geq 3$, and $V_i(x)=\sum_{j=1}^{m'}\sum_{c\in [k]} p_{c,r_j}(x)$ if $U_i=\delta_D$ and $\ell=2$ with $D=z_{r_1}\vee\cdots\vee z_{r_{m'}}$.
Further, fix $U\coloneq 8(M_{\rm all}+1)^{2\ell}$, where $M_{\rm all}\geq 1$ is the maximum fan-in of any disjunction and conjunction in $F$.
Note that $\max(0,y)$ can be computed in the first hidden layer and propagated through the subsequent layers, so it can be used in the last hidden layer.

For $y\leq 0$, we have $\max(0,y)=0$ and thus $h(x,y)=0$.
For $y>0$, we have
\[
h(x,y)=\sum_{i=1}^m (\max(0,V_i(x)-T)-\max(0,V_i(x)-T-Uy))=\sum_{i=1}^m \min(\max(0,V_i(x)-T), Uy).
\]

First, suppose that we have a yes-instance and let $x^*\in[n]^k$ correspond to a satisfying assignment.
Therefore, we have $U_i(x^*)=1$ and thus also $V_i(x^*)\geq 1$ for all $i\in [m]$.
Hence, there exists an open neighborhood $B\subseteq \R^k$ around $x^*$ with $V_i(x)>1-\frac{\varepsilon}{2}$ for all $x\in B,\, i\in [m]$.
Choosing $y^*=\frac{\varepsilon}{4U}>0$, we have for every $x\in B$
\[
V_i(x)-T-Uy^*>1-\frac{\varepsilon}{2}-(1-\varepsilon)-\frac{\varepsilon}{4}=\frac{\varepsilon}{4}>0,
\]
as $T\leq 1-\varepsilon$ holds both in the even and odd case.
Hence, there is an open neighborhood $B'\subseteq \R^{k+1}$ around $(x^*,y^*)$, such that for every $(x,y)\in B'$
\[
h(x,y)=\sum_{i=1}^m (V_i(x)-T-(V_i(x)-T-Uy))=mUy.
\]
The open set $B'$ is contained in a full-dimensional region $R^*\in \Sigma_h$, which must satisfy $h(x,y)=a_{R^*}^\top (x,y)=(mUe_y)^\top (x,y)$ for all $x\in R^*$.
Since $\|e_y\|_p=1$ for every $p\in (0,\infty]$, we have by \Cref{lem:lipschitz_region}
\[
L_p(h)=\max_{R\in \Sigma_h}\max_{\|(x,y)\|_p\leq 1}|a_R^\top (x,y)|\geq |a_{R^*}^\top e_y|=mU.
\]
Now, suppose that we have a no-instance.
For any region $R\in \Sigma_h$ with $R\cap \{(x,y): y<0\}\neq \emptyset$, we have $a_R=0$.
Thus, every region $R\in \Sigma_h$ with $a_R\neq 0$ satisfies $R\subseteq  \{(x,y): y\geq 0\}$ such that for all $(x,y)\in R$
\[
h(x,y)=\sum_{i\in I_R} (V_i(x)-T) + |J_R|Uy,
\]
where $I_R=\{i\in [m]: T<V_i(x)\leq T+Uy \text{ for all } (x,y)\in R\}$ and $J_R=\{i\in [m]: V_i(x)> T+Uy \text{ for all } (x,y)\in R\}$.
In a no-instance, there is for every $x\in \R^k$ at least one $i\in[m]$ with $V_i(x)\leq T$.
Therefore, for every $p\in [1,\infty]$, we have by the triangle inequality
\[
\|a_R\|_p=\|\nabla h_{\mid R}\|_p
\leq \sum_{i\in I_R} \|\nabla (V_i)_{\mid R}\|_p + |J_R|\cdot U\|e_y\|_p
\leq |I_R| \cdot U + |J_R|\cdot U
\leq (m-1)U,
\]
since $\|\nabla (V_i)_{\mid R}\|_p\leq U=8(M_{\rm all}+1)^{2\ell}$ holds, which follows from $L_p(s_{c,c'})\leq 2$, $L_p(p_{c})\leq 8$, and the fact that in each hidden layer, each $\gamma_C$ or $\delta_D$ function increases the Lipschitz constant by at most $\frac{1}{\varepsilon}M_{\rm all}=(M+1)M_{\rm all}\leq (M_{\rm all}+1)^2$.

Since for every $p\in(0,\infty]$, we have $\max_{\|(x,y)\|_p\leq 1}|a_R^\top (x,y)|=\|a_R\|_q$ for some $q\in [1,\infty]$ (see \Cref{lem:lipschitz_region}), it follows that 
\[
L_p(h)
=\max_{R\in \Sigma_h}\max_{\|(x,y)\|_p\leq 1}|a_R^\top (x,y)|
=\max_{R\in \Sigma_h}\|a_R\|_q
\]
which implies a separation for the $L_p$-Lipschitz constant $L_p(h)$ depending on whether the instance is a yes- or no-instance.
With this, the \textsc{$(\ell+1)$-Layer ReLU $L_p$-Lipschitz Constant} instance consisting of $L=mU$ and the underlying network of $h$ is a yes-instance if and only if $F$ is a yes-instance of \textsc{Weighted (Anti)Monotone $t$-Normalized SAT}.
\end{proof}

On the positive side, we show that for a special subclass of ReLU networks, computing the $L_1$- and $L_\infty$-Lipschitz constant is tractable.

\subsection{Tractability for Input Convex Neural Networks}
\label{subsection:icnns}

A 2-maxout network is \emph{input-convex} (ICNN) if the weight matrices of all but the first layer have only nonnegative entries, resulting in a convex function  $f(x) = \max\{a_1^\top x +b_1, \dots, a_k^\top x +b_k\}$.
The $L_p$-Lipschitz constant of $f$ is given by the maximum, taken over all linear regions $C$ of $f$, of the $L_p$-Lipschitz constant of $f$ restricted to $C$, where $f(x) = a_C^\top x + b_C$ for all $x \in C$.
Using \Cref{lem:lipschitz_region}, we obtain $L_p(f)=\max_i\|a_i\|_q$, where $\|\cdot\|_q$ is the dual norm of the $L_p$-norm.
Note that the function $f$ of the ICNN has the same $L_p$-Lipschitz constant as the function $g(x) = \max\{a_1^\top x, \dots, a_k^\top x\}$ computed by the same network where all biases are set to $0$, which implies that we might assume without loss of generality that the network does not have biases and hence computes a function $f$ that is convex and positively homogeneous.

\citet{HL24} showed that there is a small \emph{extended formulation} of $\Newt(f)$ for a function~$f$ computed by an ICNN without biases. More precisely, their proofs reveal that for a function $f\colon \R^d \to \R$ computed by an ICNN, there is a polytope $
Q \subseteq \mathbb{R}^{d+m}$ and a projection $\pi \colon \mathbb{R}^{d+m} \to \mathbb{R}^d$ such that $\pi(Q) = \Newt(f)$ and the encoding size of $(Q,\pi)$ is polynomial in the encoding size of $f$, where $Q$ is given in half-space representation. Using this, we prove the following proposition. 

\begin{proposition}\label{prop:icnn_polyhedral_lp_constants}
    Let $f \colon \R^d \to \R$ be an ICNN with encoding size~$N$. Then $L_1(f)$ can be computed in $\poly(N)$ time and $L_\infty(f)$ can be computed in $2^d\poly(N)$ time.
\end{proposition}
\begin{proof}
 By the discussion above, we can assume without loss of generality that there are no biases and $f$ is positively homogeneous. In this case, the definition of the support function implies that $L_p(f) = \max_{y \in \Newt(f)}\| y\|_q$. By \citet{HL24}, there exists a polytope $Q$ and a projection $\pi$ with $\poly(N)$ encoding size such that $L_p(f) = \max_{y \in Q} \|\pi(y)\|_q$. For $p=\infty$ and $p=1$, this maximization can be reduced to finitely many LPs: Indeed, 
$\max_{y \in Q} \|\pi(y)\|_\infty = \max_{c \in \{\pm e_1,\dots,\pm e_d\}} \max_{y \in Q} c^\top \pi(y)$, which requires solving only $2d$ LPs, while 
$\max_{y \in Q} \|\pi(y)\|_1 = \max_{c \in \{\pm 1\}^d} \max_{y \in Q} c^\top \pi(y)$, which requires solving $2^d$ LPs.  Since LPs can be solved in polynomial time, the statements follow.
\end{proof}

\section{Norm Maximization on Zonotopes}\label{sec:norm_max_fpt_approx}

In this section, we describe a connection between Lipschitz constants of neural networks and norm maximization on zonotopes.
For 2-layer ICNNs $f\colon\R^d\to \R$, we can restrict ourselves without loss of generality to the case where all output weights are equal to 1.
In this case, the Newton polytope $\Newt(f)$ is a zonotope and computing $L_p(f)$ is equivalent to maximizing the dual norm of the $L_p$-norm over this zonotope (see \citet{FGHS25}).

\citet{BP07} showed that maximizing the $L_\infty$-norm on zonotopes is solvable in polynomial time and maximizing the $L_1$-norm on zonotopes is fixed-parameter tractable for~$d$.
Our \Cref{prop:icnn_polyhedral_lp_constants} generalizes these results to $L_1$- and $L_\infty$-norm maximization to Newton polytopes of maxout ICNNs with unbounded depth (that are not necessarily zonotopes).
Note that \Cref{thm:zonotope-cont-hardness} implies that maximizing a zonotope-norm over a zonotope is W[1]-hard with respect to the dimension~$d$ (since zonotope containment is equivalent to this problem~\citep{KULMBURG202184}).
\citet{Shenmaier18} proved NP-hardness and inapproximability for $L_p$-maximization on zonotopes for $p\in [1,\infty)$ and showed a randomized (sampling based) $(1-\varepsilon)$-approximation with probability $1-1/e$ in time~$(1+2/\varepsilon)^d\poly(d,n)$ for every $\varepsilon\in(0,1)$ and an arbitrary norm.
We show that known results from \emph{subspace embedding} theory can also be used to obtain randomized approximations, which is an interesting application of these results.
The worst-case running time, however, is worse, but in practice the actual running time might still be faster.
\citet{BRR23} observed that results for $\ell_1$ subspace embeddings~\citep{CP15} yield \emph{zonotope order reductions} (or \emph{zonotope sparsifications}), that is, approximations of zonotopes with few generators. More precisely, the following can be derived.

\begin{theorem}\label{thm:order-reduction}
There is a polynomial-time algorithm which, given a matrix~$A\in\R^{d\times n}$ and $\varepsilon >0$, outputs a matrix~$A'\in\R^{d\times r}$ with~$r\in \mathcal{O}(d\log d\varepsilon^{-2})$ such that with high probability \[(1+\varepsilon)^{-1}Z(A')\subseteq Z(A)\subseteq (1+\varepsilon)Z(A').\]
\end{theorem}

\begin{proof}
    Let $A=(a_1,\dots,a_n)\in \R^{d\times n}$ be a matrix and let $Z(A)=\sum_{i=1}^n\conv\{0,a_i\}\subset \R^d$ be the corresponding zonotope.
    Defining $c\coloneq \frac{1}{2}\sum_{i=1}^n a_i$ as the center of the zonotope, we have
    \[
    Z(A)-c
    =\sum_{i=1}^n \conv\left\{-\frac{a_i}{2},\frac{a_i}{2}\right\}
    =\sum_{i=1}^n \conv\left\{0,-\frac{a_i}{2}\right\}+\sum_{i=1}^n \conv\left\{0,\frac{a_i}{2}\right\}\subset \R^d.
    \]
    We now construct the matrix $B=\left(\frac{a_1}{2},\dots,\frac{a_n}{2}\right)^\top\in \R^{n\times d}$.
    Then,
    \[
    \|Bx\|_1=\sum_{i=1}^n \left|\frac{a_i^\top x}{2}\right|=\sum_{i=1}^n \max\left\{0,-\frac{a_i^\top x}{2}\right\}+ \sum_{i=1}^n \max\left\{0,\frac{a_i^\top x}{2}\right\}
    \]
    is the support function of the zonotope $Z(A)-c$.

    Applying the polynomial algorithm of \citet{CP15}, we find a matrix $B'=(b_1',\dots, b_r')^\top\in \R^{r\times d}$ with $r\in \mathcal{O}(d\log d \varepsilon^{-2})$ such that with high probability,
    $(1+\varepsilon)^{-1}\|Bx\|_1\leq \|B'x\|_1\leq (1+\varepsilon)\|Bx\|_1$ holds for all $x\in \R^d$.
    From the duality between zonotopes and their support function, this implies $(1+\varepsilon)^{-1}Z((B'^\top,-B'^\top))\subseteq Z(A)-c\subseteq (1+\varepsilon)Z((B'^\top,-B'^\top))$.
    With $A'=(2b_1'+c,\dots, 2b_r'+c)\in \R^{d\times r}$ and $(1+\varepsilon)^{-1}\|Bx\|_1\leq \|B'x\|_1\leq (1+\varepsilon)\|Bx\|_1$, it follows that $(1+\varepsilon)^{-1}Z(A')\subseteq Z(A)\subseteq (1+\varepsilon)Z(A')$, which implies the theorem.
\end{proof}

This order reduction yields a simple randomized approximation algorithm.

\begin{theorem}
Let $\Vert\cdot\Vert$ be any norm on~$\R^d$ (computable in time~$T$). There is a randomized algorithm which, given a matrix $A\in\R^{d\times n}$ and $\varepsilon>0$, outputs a value~$\alpha\in\R$ in $\mathcal{O}((cd\log d/\varepsilon^2)^{d-1}\cdot T+\poly(n))$ time (for some constant~$c>0$) such that with high probability \[(1+\varepsilon)^{-1}\alpha\le\max_{x\in Z(A)}\Vert x\Vert\le(1+\varepsilon)\alpha.\]
\end{theorem}
\begin{proof}
Note that every norm is convex and convex functions attain their maximum on a polytope at a vertex.
On input $(A,\varepsilon)$, we run the algorithm from \Cref{thm:order-reduction} to obtain a matrix~$A'$ with $r\in \mathcal{O}(d\log d\varepsilon^{-2})$ columns in polynomial time.
The zonotope $Z(A')$ has at most~$\mathcal{O}(r^{d-1})$ vertices~\citep{Zas75},  which can be enumerated in $\mathcal{O}(r^{d-1})$ time~\citep{FFL05}.
We simply return the maximum $\Vert \cdot \Vert$-value $\alpha$ of these vertices. Then, with high probability, it holds $(1+\varepsilon)^{-1}Z(A')\subseteq Z(A)\subseteq (1+\varepsilon)Z(A')$, which implies
\[\max_{x\in (1+\varepsilon)^{-1}Z(A')}\Vert x\Vert=(1+\varepsilon)^{-1}\alpha\le \max_{x\in Z(A)}\Vert x\Vert\le (1+\varepsilon)\alpha =\max_{x\in(1+\varepsilon)Z(A')}\Vert x\Vert,\]
due to absolute homogeneity of norms.
\end{proof}
Until recently, it was open whether $L_p$-maximization on zonotopes for $p\in(1,\infty)$ is fixed-parameter tractable for~$d$~\citep{Shenmaier18,FGHS25}.
Several independent works recently resolved this question by showing W[1]-hardness for $p=2$ \citep{dewasurendra2026convex} and, more generally, for arbitrary fixed rational $p\in (1,\infty)$, also excluding algorithms running in time $\rho(d)N^{o(d)}$ assuming the ETH \citep{cao2026ellpnorm,das2026parameterized}.
Moreover, \citet{cao2026ellpnorm} present a deterministic $(1-\varepsilon)$-approximation running in time $\rho(d)\varepsilon^{-(d-1)/2}\poly(N, \log(1/\varepsilon))$ for every $\varepsilon>0$, and rule out algorithms with $(1/\varepsilon)^{o(d)}$ dependence assuming the ETH.

\section{Para-NP-hardness for Unbounded Depth}\label{sec:paraNP}
\Cref{thm:Wl-hardness} shows that the positivity problem with parameter~$d$ becomes hard for higher levels of the W-hierarchy with increasing number~$\ell$ of layers. To settle the complexity with respect to the parameter~$d$, we show NP-hardness for~$d=1$ (that is, para-NP-hardness w.r.t.\ $d$) if the number of layers is unbounded. That is, for unbounded depth, \textsc{ReLU Positivity} is not even contained in XP for the parameter~$d$ unless P $=$ NP.
Moreover, for the parameter~$\ell$, \textsc{ReLU Positivity} is hard for every finite level of the W-hierarchy (even for~$d=1$).

\begin{theorem}\label{thm:fixed_d=1_hardness_positivity}
    \textsc{ReLU Positivity} with input dimension $d=1$ is NP-hard, \textnormal{W[$t$]}-hard with respect to the number~$\ell$ of layers for every $t\ge 1$, and not solvable in~$\rho(\ell)N^{o(\ell)}$ time (where~$N$ is the input bit-length) for any function~$\rho$ assuming the ETH.
\end{theorem}

\begin{figure}
\centering
\includegraphics[width=0.8\textwidth, page=19]{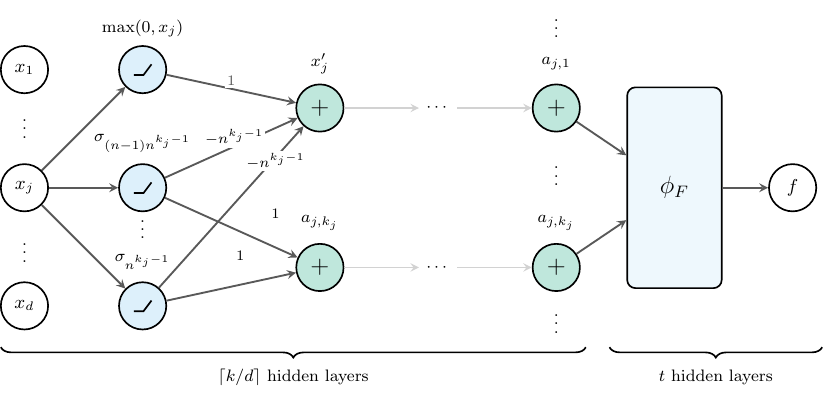}
\caption{
Illustration of the ReLU network from the proof of \Cref{thm:fixed_d=1_hardness_positivity}.
Only the subnetwork starting from $x_j$ is shown.
The first $\lceil k/d\rceil$ hidden layers successively compute the digits $a_{j,1},\dots,a_{j,k_j}$ for all $j\in [d]$.
The digits are then passed to the network $\phi_F$ with $t$ hidden layers.
The $+$ nodes represent affine transformations and are not ReLU layers themselves.
Each $\sigma$ term actually consists of two ReLU neurons, but is displayed as a single ReLU neuron for simplicity.
}
\label{fig:base_n_decomposition}
\end{figure}

\begin{proof}
Let~$t > 1$ (clearly, this implies hardness for $t=1$).
We give a polynomial-time reduction from \textsc{Weighted (Anti)Monotone $t$-Normalized SAT} to \textsc{ReLU Positivity}.
For a given instance~$(F,k)$ with~$n\ge k\ge 2$ variables $z_0,\ldots,z_{n-1}$, we actually show how to construct, for each~$d\in[k]$, a ReLU network~$f\colon\R^d\to\R$ with~$\ell=\lceil k/d\rceil+t+1$ layers which has a positive output if and only if~$F$ has a weight-$k$ satisfying assignment (the case~$d=1$ and~$\ell=k+t+1\in \mathcal O(k)$ yields the statement above).

To this end, let $r\coloneqq \lceil k/d\rceil$ and let $k_1+k_2+\cdots+k_d=k$ be an integer partition of $k$ with $k_j\leq r$. 
The idea is that the first~$r$ hidden layers compute the $k_j$-digit representation of the input number~$x_j$ with respect to the base~$n$ in parallel for all $j\in [d]$.
That is, for~$x_j\in\{0,1,\ldots,n^{k_j}-1\}$, the network computes integers~$a_{j,1},\ldots,a_{j,k_j}\in\{0,\ldots,n-1\}$ such that $x_j=\sum_{i=1}^{k_j}a_{j,i}n^{i-1}$.
The~$k$ numbers~$a_{1,1},\ldots,a_{1,k_1},a_{2,1},\ldots,a_{2,k_2},\ldots,a_{d,k_d}$ then serve as inputs for a ReLU network~$\phi_F\colon\mathbb{R}^k\to\mathbb{R}$ with $t+1$ layers which simulates the formula~$F$.
Note that we already constructed such a network in the proof of \Cref{thm:Wl-hardness}.

Hence, in the following, we construct a network $\phi_j\colon\R\to\R^{k_j}$ computing~$a_{j,1},\ldots,a_{j,k_j}$.
The network~$\phi_j$ first computes~$a_{j,k_j}$. Note that $a_{j,k_j}=\max\{a\mid 0\le a < n, an^{k_j-1} \le x_j\}$.
The first hidden layer contains two neurons for each~$a\in[n-1]$ approximating a ``step function''
\[\sigma_{an^{k_j-1}}(x_j)\coloneqq\max(0,2(x_j-an^{k_j-1}+1/2))-\max(0,2(x_j-an^{k_j-1})).\]
Note that $\sigma_{an^{k_j-1}}(x_j)=1\implies x_j \ge an^{k_j-1}$ and $x_j\le an^{k_j-1}-1\implies \sigma_{\ge an^{k_j-1}}(x_j)=0$.
The affine part of the second layer then computes the value~$a_{j,k_j}$ by summing up all step functions from the previous layer
\[a_{j,k_j} = \sum_{a\in[n-1]}\sigma_{an^{k_j-1}}(x_j).\]
Moreover, it analogously computes the step functions $\sigma_{an^{k_j-2}}(x_j')$ (where~$x_j'\coloneqq x_j- a_{j,k_j} n^{k_j-1}$) which can be summed up to the value $a_{j,k_j-1}$ in the next layer.
Note that an additional neuron computing $\max(0,x_j)$ is necessary to compute $x_j'$.
Further, we can use one extra neuron per hidden layer to propagate the computed $a_{j,i}$ values to the output layer.
Continuing this construction, we end up with a network with $k_j$ hidden layers which on input~$x_j=\sum_{i=1}^{k_j}a_{j,i}n^{i-1}\in\{0,\ldots,n^{k_j}-1\}$ outputs~$a_{j,k_j},\ldots,a_{j,1}$ (as can easily be verified).

Stacking this construction for all $j\in [d]$ yields a network with at most $\max_{j\in [d]}k_j\leq r$ hidden layers.
Note that not all sub-networks necessarily have the same number of hidden layers.
To make the number of hidden layers the same in each sub-network, whenever $k_j< \max_{j^*\in [d]}k_{j^*}$ for a $j\in [d]$, we can repeatedly add $k_j$ neurons that compute $\max(0, a_{j,i})=a_{j,i}$ for all $i\in [k_j]$ after the last hidden layer of the sub-network of $x_j$.
As already mentioned above, we finish the construction of~$f$ by appending the network~$\phi_F$ (see \Cref{fig:base_n_decomposition} for an illustration).

For the correctness, assume that~$F$ is satisfied by setting variables~$z_{i_1},\ldots,z_{i_k}$ to true.
Then let~$x_j^*\coloneqq \sum_{q=1}^{k_j}i_{\left(q+\sum_{h=1}^{j-1} k_h\right)}n^{q-1}$ for all $j\in [d]$.
Clearly, we have $(\phi_{1}(x_1^*),\dots, \phi_{d}(x_d^*))=(i_1,\ldots,i_k)$ and since~$F$ is satisfied, it follows (by the proof of \Cref{thm:Wl-hardness}) that
\[f(x^*_1,\ldots,x_d^*)=\phi_F(\phi_{1}(x_1^*),\dots, \phi_{d}(x_d^*))=\phi_F(i_1,\ldots,i_k)>0.\]

Conversely, if~$(F,k)$ is a no-instance, then~$\phi_F(x)\le 0$ for all~$x\in\mathbb{R}^k$ (by \Cref{thm:Wl-hardness}). Hence, $f(x)\le 0$ for all $x\in\mathbb{R}^d$.
\end{proof}

Analogously to the proofs of the previous sections, we obtain the following corollary.
\begin{corollary}\label{cor:fixed_d=1_hardness_approx_lipschitz}
    The following ReLU network problems with~$d=1$ are NP-hard, \textnormal{W[$t$]}-hard with respect to the number~$\ell$ of layers for every $t\ge 1$, and not solvable in~$\rho(\ell)N^{o(\ell)}$ time (where~$N$ is the input bit-length) for any function~$\rho$ assuming the ETH:
    \begin{itemize}
        \item Deciding whether a ReLU network computes a non-zero function,
        \item approximating the maximum of a ReLU network within any multiplicative factor,
        \item for every $p\in(0,\infty]$, approximating the $L_p$-Lipschitz constant of a ReLU network within any multiplicative factor.
    \end{itemize}
\end{corollary}
For surjectivity, we obtain hardness for $d=2$, in sharp contrast to the $d=1$ case, where surjectivity is solvable in polynomial time (see \Cref{thm:WP-containment_param_layers}).
\begin{corollary}\label{cor:fixed_d=2_hardness_surjectivity}
    \textsc{ReLU Surjectivity} with input dimension $d=2$ is NP-hard, \textnormal{W[$t$]}-hard with respect to the number~$\ell$ of layers for every $t\ge 1$, and not solvable in~$\rho(\ell)N^{o(\ell)}$ time (where~$N$ is the input size) for any function~$\rho$ assuming the ETH.
\end{corollary}
\begin{proof}
    Let $f\colon\R\to\R$ be the network constructed in the proof of \Cref{thm:fixed_d=1_hardness_positivity} with $d=1$.
    We replace the neuron computing $\max(0,x)$ in the first hidden layer by three neurons computing $\max(0,x)-\max(0,2(x-n^k))+\max(0,x-2n^k)$, which is $\max(0,x)$ for $x\leq n^k$, then decreases to $0$ with slope $-2$, and is identically~$0$ for all $x\geq 2n^k$.
    Further, we replace $\sigma_{an^{k-1}}$ in the first hidden layer by
    \[
    \sigma'_{an^{k-1}}(x)
    =\max(0,2(x-an^{k-1}+1/2))-\max(0,2(x-an^{k-1}))-\max(0,2(x-n^k))+\max(0,2(x-n^k-1/2)),
    \]
    which is $\sigma_{an^{k-1}}(x)$ for $x\leq n^k$, then decreases to $0$ with slope $-2$, and is identically~$0$ for all $x\geq n^k+1/2$.
    Note that the modified network still has a positive point if and only if $f$ has a positive point.
    This modified network now satisfies the assumption of \Cref{lem:homogenization_function}.
    Further, its homogenization $h\colon\R^2\to \R$ satisfies $h(x,0)=0$ for all $x\in \R$, since the deletion of biases leads to a cancellation of all terms in the second hidden layer.
    By the same argument as in the proof of \Cref{thm:positivity_hardness_without_bias}, $h$ has a positive point if and only if $f$ has a positive point.
    Moreover, $h(0,1)<0$, which implies that $h$ is surjective if and only if it has a positive point.
\end{proof}

We close with an in-depth discussion of the running time lower bounds implied by our results above.
For a function $g\colon\N_{\geq 2}\times \N_{\geq 1}\to \R_{>0}$, we write $g(\ell,d)\in o((\ell-1) d)$ if, for every sequence $\left((\ell_i, d_i)\right)_{i\in \N}$ with $(\ell_i, d_i)\in \N_{\geq 2}\times \N_{\geq 1}$ and $(\ell_i -1)d_i\to \infty$, we have $\lim_{i\to \infty }\frac{g(d_i,\ell_i)}{(\ell_i-1) d_i}=0$.
Under this definition, the ETH-based lower bounds for constant $d$ and constant~$\ell$ already imply that \textsc{ReLU Positivity} cannot be solved in $\rho(\ell,d)N^{o((\ell-1) d)}$ time for any computable function $\rho$ assuming the ETH, as restricting such an algorithm to $\ell=2$ would give a running time of $\rho(2,d)N^{o(d)}$, contradicting the ETH-based lower bound for the 2-layer case.
However, this does not exclude running times whose exponent is linear in $d$ and $\ell$ separately.
For example, $\ell+d \notin o((\ell-1) d)$ under the above definition, since for the sequence $\ell_i=2,\,d_i=i$, we have $\lim_{i\to \infty }\frac{\ell_i+d_i}{(\ell_i-1) d_i}=1$.
Thus, the above lower bound does not exclude an algorithm with running time $N^{\mathcal{O}(\ell+d)}$.
However, the reduction in the proof of \Cref{thm:fixed_d=1_hardness_positivity} allows us to exclude algorithms with such runtimes.
\begin{corollary}\label{cor:eth_lb_combined}
    \textsc{ReLU Positivity} cannot be solved in time $\rho(\ell,d)N^{o((\ell+d)^2)}$ (where $N$ is the input bit-length) for any computable function $\rho$ assuming the ETH.
\end{corollary}
\begin{proof}
    Suppose that \textsc{ReLU Positivity} can be solved in $\rho(\ell,d)N^{g(\ell,d)}$ time for some computable function $\rho$ with $g(\ell,d)\in o((d+\ell)^2)$.
    Fix a $t>1$ and choose $d=\lceil k^{1/2}\rceil$ for the network constructed in the proof of \Cref{thm:fixed_d=1_hardness_positivity}.
    Then, we have $\ell=\lceil k/d\rceil+t+1\in\mathcal O(k^{1/2})$ and $(\ell+d)^2\in\Theta(k)$.
    On these instances, we have $g(\ell,d)\in o(k)$, which would imply that \textsc{Weighted (Anti)Monotone $t$-Normalized SAT} can be solved in $\rho'(k)n^{o(k)}$ time for some computable function $\rho'$, which contradicts the ETH.
\end{proof}
In particular, \Cref{cor:eth_lb_combined} excludes running times $N^{\mathcal{O}((\ell+d)^{2-\varepsilon})}$ for every constant $\varepsilon>0$.
Since the algorithms in \Cref{thm:enumeration_algorithms} run in $N^{(\ell-1)d+\mathcal{O}(1)}$ time and $(\ell-1)d\leq (\ell+d)^2$, the quadratic dependence on the combined parameter $\ell+d$ in the exponent is asymptotically optimal under the ETH.

\section{Conclusion}\label{sec:conclusion}

We proved the strongest computational hardness results for fundamental ReLU network verification problems known so far.
Note that several considered problems can be phrased in terms of maximizing a certain norm over a zonotope; a problem with numerous applications in other areas.
Most importantly, our results imply that simple ``brute-force'' enumeration algorithms are basically best possible with respect to the dependency of the running time on the input dimension.
Thus, we settled the parameterized complexity of a wide range of problems almost completely.

Moreover, our results show that even for a single hidden layer it does not help to assume that the weights are sparse and small, since our 2-layer constructions use only a constant number of (polynomially bounded) non-zero weights for each hidden ReLU neuron.
It is thus not easy to formulate a general guidance to circumvent this hardness in practice.
One would have to make very specific assumptions on the network structure to ensure that the number of linear regions is small and easy to enumerate. It is not clear which assumptions would be natural here and whether networks trained on real-world data satisfy them.
Alternatively, one might use techniques (possibly incorporated into the training process) that guarantee efficient verification or use special architectures (such as ICNNs).
We also discussed some tractable cases for restricted subclasses of problems as well as a randomized FPT-approximation.
Overall, our hardness results testify that such techniques and the use of heuristics are indeed required in practice to achieve reasonable running times.

An interesting question is whether running times in $\mathcal{O}(n^{cd})$ for $c<1$ can be achieved for the considered problems.
Last but not least, to fully settle the parameterized complexity, it remains to find out the precise relation between the ``ReLU-hierarchy'' and the W-hierarchy.
The central question is whether \textsc{$\ell$-Layer ReLU Positivity} parameterized by~$d$ is complete for the class W[$\ell-1$]?
Note that this would imply a ``computational'' depth separation which means that there are $\ell$-layer networks which cannot be turned in FPT-time into equivalent networks with~$\ell-1$ layers.
However, we do not even know whether the 2-layer case is contained in~W[$t$] for some~$t\ge 1$ (we only know containment in W[P]) or whether it is W[2]-hard. A first step could be to show containment in a finite level Th[$t$] (which contains W[$t$]) of the Th-hierarchy (defined via threshold gates, see \cite{PSS24}).
Notably, for $\ell=3$ layers, we do not even know whether the problem is contained in W[P].

\bibliography{ref}
\bibliographystyle{plainnat}

\end{document}